\documentclass[12pt,preprint]{aastex}
\begin{document}

\title{Comparison of 3.6 -- 8.0 Micron {\it Spitzer}/IRAC Galactic Center Survey 
Point Sources with Chandra X--Ray Point Sources in the Central 40x40 Parsecs}

\author{
R.~G.~Arendt\altaffilmark{1,2}, 
D.~Y.~Gezari\altaffilmark{3},
S.~R.~Stolovy\altaffilmark{4},
K.~Sellgren\altaffilmark{5}, 
R. Smith\altaffilmark{6},
S.~V.~Ram\'{\i}rez\altaffilmark{7}, 
F.~Yusef-Zadeh\altaffilmark{8},
C.~J.~Law\altaffilmark{8,9}, 
H.~A.~Smith\altaffilmark{10},
A.~S.~Cotera\altaffilmark{11}, 
S.~H.~Moseley\altaffilmark{12}
}

\altaffiltext{1}
{CRESST/UMBC/GSFC, Code 665, NASA/Goddard Space Flight Center,
8800 Greenbelt Road, Greenbelt, MD 20771; richard.g.arendt@nasa.gov}
\altaffiltext{2}
{Science Systems and Applications, Inc.}
\altaffiltext{3}
{NASA/Goddard Space Flight Center, Code 667, 8800 Greenbelt Road, Greenbelt, MD 20771}
\altaffiltext{4}
{{\it Spitzer} Science Center, California Institute of Technology, 
Mail Code 220-6, 1200 East California Boulevard, Pasadena, CA 91125}
\altaffiltext{5}
{Department of Astronomy, Ohio State University, 140 West 18th Av.,
Columbus, OH, 43210, USA}
\altaffiltext{6}
{NASA/Goddard Space Flight Center, Code 662, 8800 Greenbelt Road, Greenbelt, MD 20771}
\altaffiltext{7}
{IPAC, California Institute of Technology,
Mail Code 100-22, 1200 East California Boulevard, Pasadena, CA 91125}
\altaffiltext{8}
{Department of Physics and Astronomy, Northwestern University,
Evanston, IL 60208}
\altaffiltext{9}
{now at Sterrenkundig Instituut ``Anton Pannekoek'', Universiteit van Amsterdam}
\altaffiltext{10}
{Harvard-Smithsonian Center for Astrophysics, 
60 Garden Street, Cambridge, MA 02138}
\altaffiltext{11}
{SETI Institute, 515 North Whisman Road, Mountain View, CA 94043}
\altaffiltext{12}
{NASA/Goddard Space Flight Center, Code 665, 8800 Greenbelt Road, Greenbelt, MD 20771}

\begin{abstract}
We have studied the correlation between 2357 Chandra X--ray point sources in a 
$40 \times 40$ parsec field and $\sim$20,000 infrared sources we observed in the 
corresponding subset of our $2\arcdeg \times 1.4\arcdeg$ {\it Spitzer}/IRAC Galactic 
Center Survey at 3.6--8.0 $\micron$, using various spatial and X--ray hardness thresholds. 
The correlation was determined for source separations of less than 
$0\farcs5$, 1$''$ or 2$''$. Only the soft X--ray sources show any correlation
with infrared point sources on these scales, and that correlation is very weak. 
The upper limit on hard X--ray sources that have infrared counterparts 
is $<1.7\%$ (3$\sigma$). However, because of the confusion limit of the IR catalog, 
we only detect IR sources with absolute magnitudes $\lesssim 1$. As a result, a stronger 
correlation with fainter sources cannot be ruled out.
Only one compact infrared source, IRS~13, coincides with any of the dozen prominent X--ray emission features in the $3 \times 3$
parsec region centered on Sgr A*, and the diffuse X--ray and infrared 
emission around Sgr A* seems to be anti-correlated on a few-arcsecond scale. We compare 
our results with previous identifications of near--infrared companions to 
Chandra X--ray sources.
\end{abstract}


\section{Introduction}

X-ray surveys of the Galactic center with 
{\it Chandra X-Ray Observatory} (Wang et al.\ 2002; Muno et al.\ 2003) have provided a deep sampling of the 
population of X--ray point sources that shows a large increase in source density 
toward the Galactic Center. These X--ray sources have been modeled
as a population mix of various sorts of X-ray binaries, Wolf-Rayet
stars, nearby X-ray active stars in the foreground, and background AGN
(Pfahl et al.\ 2002;
Belczynski \& Taam 2004;
Ebisawa et al.\ 2005;
Ruiter et al.\ 2006;
Liu \& Li 2006;
Muno et al.\ 2006).  However, the high extinction toward the Galactic center 
prohibits the detection of visible light emitted by the stellar components in the expected 
binaries.  Near--IR searches have also had little success in detecting IR counterparts
of the X--ray point sources. Several OB stars and Wolf-Rayet stars have been identified with X-ray sources by near-IR spectroscopy (Muno et al.\ 2006, Mikles et al.\ 2006, Mauerhan et al.\ 2007), but the paucity of near-IR detections sets limits that suggest only a small fraction of the X--ray sources can be high mass X--ray binaries (HMXBs; Laycock et al.\ 2005; Bandopadhyay et al.\ 2006).

Our {\it Spitzer Space Telescope} IRAC survey of the Galactic Center
(Stolovy et al.\ 2006, S. Stolovy et al.\ 2008 in preparation) provides a new opportunity to 
search for IR counterparts to the X--ray sources. IRAC observations cover
four broad bands at 3.6, 4.5, 5.8 and 8 $\micron$. The survey imaged a 
$2.0\arcdeg \times 1.4\arcdeg$ ($280 \times 200$ parsec at 8.0 kpc)
region of the Galactic Center with a nominal resolution of $\sim2''$ (Figure 1).
Since our observations are at longer wavelengths 
than ground--based near--IR ($J, H, K$) observations, extinction should be less
of a hinderance to the detection of stellar companions. Furthermore, at the longest IRAC 
wavelengths (5.8 and especially 8 $\micron$), IRAC is sensitive to 
circumstellar dust emission which may cause significant extinction at shorter wavelengths.
Thus, comparison of the X--ray and IR point source catalogs may reveal 
stellar companions which are at an evolutionary stage where they produce 
large quantities of dust, or are simply too heavily attenuated by the line of sight
extinction at shorter wavelengths. 

We calculated the correlation between 2357 hard and soft Chandra X--ray 
sources identified and catalogued by Muno et al.\ (2003) 
and the $\sim$20,000  {\it Spitzer}/IRAC infrared point sources that lie within 
a $40 \times 40$ parsec ($20 \times 20$ arcmin) field at the Galactic Center (Figure 2). 
The IR sources are a small subset of our 
full catalog (Ram\'irez et al.\ 2008) which has a mean confusion limit of $[3.6] = 12.4$ mag. 
We divide the Chandra sources by their hardness because the high column density of gas towards the GC, $N_H$ $\sim$ 5 $\times$ 10$^{22}$ cm$^{-2}$, absorbs all soft X-rays.  Thus soft X-ray sources must
be in the foreground towards the GC; hard X-ray sources can be at
the 8 kpc distance of the GC or can be foreground/background sources. 

\section{Analysis}
\subsection{IR/X--ray Point Source Correlations}

The positional uncertainty of the IRAC point sources is correlated with wavelength. In the final band--merged catalog, the reported position
is that measured at the shortest IRAC wavelength at which each source was detected.
Within a radius of $10'$ from the galactic center 86\% of the IRAC sources are detected at 
3.6 or 4.5 $\micron$. According to Ram\'irez et al.\ (2008), 90\% of these sources
have positional errors of $<0.16''$. 
The positional uncertainties of the X--ray sources are reported by Muno et al.\ (2003)
to be increasing with distance from the center of the field. Based on the given information,
we have assigned uncertainties to the X--ray positions that are the larger of $0\farcs3$ or $0.209''\ e^{\theta/225''}$,
where $\theta$ is the distance from the center of the field. With this prescription, only 39 
of the 2357 X--ray sources have positional uncertainties $>2\farcs5$, and 1645 ($\sim70\%$)
have uncertainties $<0\farcs8$.

In light of these positional uncertainties, we searched for infrared sources 
that fell within three different limits ($0\farcs5$, 1$''$ and 2$''$) of the 
Muno et al.\ (2003) X--ray source positions. The tightest constraint here ($0\farcs5$), 
should include all
associations between infrared and X--ray sources with high positional accuracies. Some associations
between sources with larger positional errors may be missed, but the tight limit
will best exclude coincidental associations between unrelated sources along the same line of sight.
The looser limits were employed to provide a more complete census of the total number of
possible associations, and to provide a statistical test for random unrelated
associations, which should increase directly proportionally to the area of the constraint.

Figures \ref{fig_ch1x}a and \ref{fig_ch4x}a show all the X--ray source locations plotted on the 
IRAC 3.6 and 8 $\micron$ images. Figures \ref{fig_ch1x}b and \ref{fig_ch4x}b show only the 
X--ray sources with IR counterparts within the specified limits ($<2''$, 1$''$, $0\farcs5$ 
are red, green and blue symbols respectively).
Figure \ref{fig_dots}a--b illustrates the cataloged distribution of IR sources and 
X--ray sources. The distribution of IR sources appears uniform only because it is confusion 
limited. For stars much brighter than the confusion limit, the density is peaked at the 
Galactic center (Ram\'irez et al.\ 2008) as is that of the X--ray sources.

The number of X--ray sources $N$ found coincident with an IR source, for each of the 
three radial constraints and at each IRAC wavelength is listed in Table 1. 
The statistical uncertainty on $N$ is given by $\sigma_N = \sqrt{N}$.
The X--ray and infrared sources in the 
study field show an extremely weak correlation. Fewer than 7\% of the 2357 X--ray 
sources had 3.6 $\micron$ counterparts in the catalog of IRAC infrared sources 
in a 1$''$ sampling radius, and most of these are likely to be false identifications. 
The number of counterparts decreases with increasing wavelength.

\subsection{Likelihood of Chance Associations}
We determined 
the likelihood of false correlations in the crowded infrared field by repeating the 
same correlation analysis eight additional times, but with the X--ray source position 
template offset from the nominal location, in a $5''$--pitch regular grid of eight positions N--S and E--W of center.
Figure \ref{fig_dots}c shows the distribution of the X--ray sources with 3.6 $\micron$ 
counterparts (within 1$''$), and Figure \ref{fig_dots}d shows the distribution when the 
IR sources are artificially offset by 5$''$.
The mean number of X--ray sources with IR counterparts in these offset comparisons 
is listed in Table 1 as $M$, with $\sigma_M = \sqrt{M/8}$.
The number of coincidences found in excess of chance is $N-M$. This excess is 
typically 1\% of the X--ray  sources, and never larger than 3 times the statistical uncertainty.
However, the fact that $M$ increases proportionally to the area of the matching constraint,
while $N$ increases more slowly, is another indication that (depending on wavelength) 
roughly 10-30 of the matching IR and X--ray sources are real associations.
There are no cases of multiple candidates within circles of $0\farcs5$ or 1$''$,
and only 11 sources with 2 candidates within a 2$''$ circle. These numbers 
are lower than expected by random chance because the width of the IRAC beam prevents
resolving sources that are closer together than $\sim3''$.

As some of the X--ray sources do have fairly large positional uncertainties, we have repeated
the matching using only the X--ray sources with uncertainties $<0\farcs8$. These results are
shown in Table 2. Qualitatively, the results are similar to Table 1, in that the offset
matching tests usually find fewer matches than the properly aligned tests. Also the number
of matching sources ($N$) again rises more slowly than is expected (and measured by $M$)
for purely random associations. However, because the overall sample size is reduced by a
factor of 0.7, the results are not as statistically significant as those obtained using
the full sample.

An interesting difference appears if we separate the X--ray sources according to their spectral hardness.
There is a modest excess in the number of soft X--ray sources (defined as 
those with detections in the 0.5--2.0 keV band) with infrared 
counterparts above that expected from random chance (Table 3). 
There is a 4$\sigma$ excess in the number of soft X--ray sources with IR 
counterparts at 3.6 $\micron$ compared to a 2$\sigma$ excess 
in the correlations at 8.0 $\micron$. Again the trend persists, but with lower
statistical significance, if we limit the X--ray data set to those sources 
with positional uncertainties $<0\farcs8$ (Table 4).
Based on the soft X--rays observed from these sources 
(and therefore their relatively low column densities), we expect that the soft X--ray 
sources are foreground objects, and not at the distance of the Galactic Center. The 
excess soft X--ray/IR correlation is therefore likely due to these sources being 
nearby, rather than being intrinsically bright in both wavebands. 
In contrast, the hard X--ray sources seem to have only as many infrared counterparts 
as would be expected from random associations (Tables 5 and 6). This is indicated both 
by the comparison of the aligned and offset matching ($N$ vs. $M$), and by the way
$N$ increases proportionally to the area of the matching constraint.

\section{Results}
We provide a listing of IR and X--ray sources coincident within 1$''$ in Table 7. 
The table is provided to facilitate follow up research on these sources, but we
remind the reader that the majority of these associations are likely to be merely 
coincidental.
The table contains designations and coordinates from each catalog, along with the IR 
magnitudes, an indication X--ray spectral hardness, and the actual 
separation between the associated sources. Much additional information can be found 
by consulting the original catalogs (Muno et al.\ 2003; Ram\'irez et al.\ 2008).

\subsection{Nature of the Correlations}
The spatial distribution of soft X--ray sources is less concentrated 
toward the Galactic Center than that of the hard sources. Figure \ref{fig_cumul}
shows the cumulative distributions of distance from Sgr A* for all X--ray sources
and for the hard and soft sources separately. Also shown is the distribution 
of the 51 sources which had 3.6 $\micron$ IR counterparts within $0\farcs5$.
These distributions can be compared statistically using the 
Kolmogorov--Smirnov (K--S) test (e.g.\ Press et al.\ 1986). This test
measures the maximum separation $D$ between two cumulative distributions. Then we
calculate that probability of having the two distributions differ by $D$ or larger,
under the assumption that they are sampled from the same parent distribution. 
By this statistic, the probability of finding the observed $D(hard-soft)$ is 
essentially 0, if the two types of X--ray source really have the same radial 
spatial distribution with respect to Sgr A*. The probability of finding the observed 
$D(hard - 3.6\micron)$ is 0.01, and the probability of finding the observed 
$D(soft - 3.6\micron)$ is 0.72. Thus, the K-S test indicates that it is statistically 
unlikely that the hard X--ray sources and the set of matching 3.6 $\micron$ IR sources
share the same degree of clustering toward the Galactic Center. Whereas the 
distribution of the matching 3.6 $\micron$ IR associations is entirely
consistent with that of the soft X--ray sources. This is further evidence that a
substantial majority of IR counterparts are associated with the soft X--ray sources
rather than the more numerous hard sources. For comparison the distribution 
of all 3.6 $\micron$ sources within the same area of the X-ray sources is also 
shown in Figure \ref{fig_cumul}. As would be inferred from Figure \ref{fig_dots}a, this 
distribution is very close to (but not exactly) uniform. The probability of finding 
$D(uniform - all\ 3.6\micron)$ is 0.03.

\subsection{Infrared Colors of Counterparts to X--ray Sources}

We next investigated the colors of the potential IR counterparts, to see if they might 
be useful for separating intrinsic IR/X--ray sources from the random associations. The 
color--magnitude diagram for the 17613 IRAC sources in the $40 \times 40$ parsec study 
region that were detected at both 3.6 and 4.5 $\micron$ is shown in Figure \ref{fig_cmd}. 
The circles and squares indicate IR stars that are identified with hard and soft X--ray sources, respectively, in the Muno et al.\ (2003) catalog. 42 X--ray sources had IRAC sources falling within $0\farcs5$ of the nominal X--ray positions (large circles/squares), 130 had IRAC sources 
within 1$''$ (medium circles/squares), and 394 had IRAC sources falling within 2$''$ (small circles/squares). See Tables 1--3 for details, including statistics for hard and soft X--ray sources. 

The colors of the possible infrared counterparts do not seem to be unusual in any way. 
The distribution of colors of those infrared stars that have coincident X--ray sources 
is essentially the same as the distribution of all other infrared stars in our sample. 
There is also no apparent differentiation between the colors of infrared stars as a 
function of their distance from the X--ray source.
The K-S test reveals no significant difference between the color distributions of 
the infrared candidates and the larger sample of all infrared sources within the field. 
Thus, the observed colors of the IRAC point sources are of little use in attempting to 
discriminate between real and coincidental X--ray/IR correlations.

\subsection{Infrared and X--ray Sources at the Central Parsec}

Casual inspection of the Chandra and {\it Spitzer}/IRAC images gives the impression 
that the bright X--ray cluster at the Galactic Center is well-correlated with the 
infrared emission.  However, closer examination
shows that only one of the bright, compact mid-infrared
sources, IRS 13, coincides with any of the discrete X-ray sources in the
central 3 $\times$ 3 parsecs (Figure \ref{fig_zoom}). 
The separation between the X--ray and IRAC positions is $<1''$
(see source SSTGC 0524504 in Table 7).
Notably, IRS 13 has been proposed as a candidate host for an intermediate black hole (Maillard et al.\ 2004). While we do also detect IRS 10* (= IRS 10EE), the mid-infrared counterpart to an SiO maser source, with Spitzer/IRAC, and Peeples et al.\ (2007)
identify a Chandra X-ray source with IRS 10*, that X-ray source is more than 2$''$ east of IRS 10* so it does not meet our correlation criteria.
Peeples et al.\ (2007) find
spatial coincidences between Chandra X-ray sources and
five other point sources they observe at 1.6 and 2.2 $\micron$,
but we do not detect any of these sources with $Spitzer$/IRAC.
Thus, with the exception of IRS 13, none of the
point sources in the
central 3 $\times$ 3 parsecs imaged in the $Spitzer$/IRAC 3.6 -- 8.0 $\micron$ survey, and none of the compact 12 $\micron$ IRS sources imaged with
sub-arcsec resolution by Gezari et al.\ (1996), have X-ray counterparts.  The 
X--ray emission in the immediate vicinity of Sgr A* is discussed in Gezari et al.\ (2006).

\section{Discussion}

Several population studies have been done to estimate
what the relative contributions of different X-ray sources
are to the Galactic Center catalog of Muno et al.\ (2003)
or other Galactic Center X-ray surveys.
These models give insight into what infrared counterparts
we might hope to find with Spitzer/IRAC.
The Galactic Center is thought to contain a mix of
CVs, magnetically accreting CVs (intermediate polars),
low-mass X-ray binaries (LMXBs), high-mass X-ray binaries (HMXBs), massive stars with strong winds, colliding wind binaries, and pulsars
(Pfahl et al.\ 2002; Belczynski \& Taam 2004; Ruiter et al.\ 2006;
Liu \& Li 2006; Muno et al.\ 2006).
We would like to know which of these are most likely to be
detectable at a distance of 8 kpc with Spitzer/IRAC.
The donor stars for HMXBs, massive stars with strong winds,
and colliding wind binaries are O stars and Wolf-Rayet stars.
These stars are luminous enough to be detected with Spitzer/IRAC, especially if they have
circumstellar dust or heat the surrounding interstellar medium.
The donor stars for CVs, intermediate polars, and LMXBs
are probably not luminous enough to be detected with Spitzer/IRAC
unless they are a red giant or they are surrounded by circumstellar dust.

The INTEGRAL gamma--ray observatory has discovered a new class of hard X--ray
source in the Galactic plane: highly obscured HMXBs, with hydrogen column
densities much higher than predicted by their extinction, due to intrinsic
absorption in the wind of the supergiant secondary (Dean et al.\ 2005; Walter et al.\ 2006).
{\it Spitzer}/IRAC photometry of highly obscured supergiant HMXBs has detected
moderate infrared excesses from a hot dust component, but with roughly a 25\% success rate
(Kaplan et al.\ 2006; Rahoui et al.\ 2008). However, none of these cases show 
evidence for a large mass of cooler dust in the systems, which could lead to a 
rising mid-IR spectrum at wavelengths $\gtrsim 5\micron$. Our lack of detection of a 
large number of 8 $\micron$ counterparts suggests that the GC X-ray sources 
also lack massive dusty shells. 

Pfahl et al.\ (2002) conclude that most of the X-ray sources found by Wang et al.\ (2002),
in a wider but shallower Chandra survey of the Galactic Center,
are HMXBs: neutron stars with O or B star companions.
Other authors, however, conclude that only a small
fraction of the Muno et al.\ (2003)
Galactic Center X-ray sources are associated with massive stars
(Belczynski \& Taam 2004; Ruiter et al.\ 2006;
Liu \& Li 2006; Muno et al.\ 2006).
Muno et al.\ (2006), Mikles et al.\ (2006), and Mauerhan et al.\ (2007)
have combined Chandra point source catalogs of the Galactic
Center with the 2MASS catalog to find candidate massive stars.
Muno et al.\ (2006) also includes 3.6 cm radio data.
They discard all soft X-ray sources, because the interstellar
absorption to the Galactic Center will absorb all soft
X-ray emission from there.
They also discard all 2MASS sources with blue J-K colors
as being foreground sources, because the interstellar extinction
to the Galactic Center reddens all sources there.
They use near-IR spectra to identify three (Muno et al.\ 2006),
one (Mikles et al.\ 2006), and two (Mauerhan et al.\ 2007)
massive stars in the Galactic Center with X-ray emission.
That so much effort has gone into identifying a total of six
massive stars with X-ray emission, either from colliding
winds or from accretion onto a compact companion,
out of more than a thousand hard X-ray sources in the
Galactic Center, underscores the difficulty of
observing infrared emission from the donor stars for X-ray sources in the Galactic Center. 

Bandyopadhyay et al.\ (2005), reported that roughly 75\% of a sample of 77 hard X--ray 
sources had faint ($K = 13 - 20$) candidate K--band counterparts within a $1\farcs3$ radius. 
However, they 
noted that this is exactly the number of random associations that they predict from a 
Monte Carlo simulation. Thus although Bandyopadhyay et al.\ (2005) identified 58 K--band ``candidates'' 
none of these sources are likely to be real infrared counterparts to the hard X--ray sources 
in their sample. Correlations at H and J bands did show small statistical excess 
above the randomly expected number of associations. 
Follow--up spectroscopic observations by Bandyopadhyay et al.\ (2006) failed to find 
indications of accretion (e.g.\ Br $\gamma$ emission) among 28 candidates observed. 
They conclude that the apparent associations are merely foreground stars, and 
that the X--ray sources are likely dominated by a population of low mass X--ray 
binaries (LMXBs) and cataclysmic variables (CVs) with $K > 20$.  
Laycock et al.\ (2005) compared 1453 hard and 105 soft X--ray sources with near infrared stars from 
the 2MASS survey and concluded that high--mass X--ray binaries are not the dominant 
hard X--ray source population. 

Unlike the hard X--ray sources examined by Bandyopadhyay et al.\ (2006), we clearly do detect (if only in a statistical sense) a population of sources with both soft X--ray and infrared emission, notably in the 3.6 and 4.5 $\micron$ bands (a net excess of $\sim34\pm8$ sources in these bands when a positional agreement of $< 1''$ is required; see Table 3). These sources represent $\sim6\%$ of the complete soft X--ray sample and, as noted previously, are likely foreground sources rather than being intrinsically bright in IR and X--rays.  
Ebisawa et al.\ (2005) observed a blank part of the galactic
plane with Chandra, and were able to find 2MASS identifications
for almost all their soft X-ray sources; they concluded they
were nearby active stars on the main sequence.
For comparison, Sidoli et al.\ (2001) found 107 sources in a 12 deg$^2$ region around the GC surveyed with the ROSAT PSPC, which was sensitive to X--rays between 0.1--2.4 keV.  Of these 107, they identified 20 (or 19\%) as being associated with stars, noting that these have softer or less absorbed spectra.  Although our data cannot conclusively determine if the softer X--ray sources with IR counterparts are indeed from a different population than the sources without counterparts, this could be tested with more observations and may represent a method to eliminate foreground sources from a GC catalog.

The donor stars in both LMXBs and CVs at the GC are expected to be too faint to be 
detected in our IRAC survey. Over the full IRAC GC survey, Ram\'irez et al.\ (2008)
set mean confusion limits at 12.4, 12.1, 11.7 and 11.2 mag for 3.6, 4.5, 5.8 and 8 $\micron$
respectively. However, because of the general increase in the volume density of stars with decreasing 
distance from the Galactic Center, confusion limits are about 2 magnitudes brighter in the 
vicinity of Sgr A (see Fig. 8 of Ram\'irez et al.\ 2008). Given a distance modulus of 7.26 and 
extinction of about 1.5 - 2 magnitudes, scaled from $A_K \approx 3.3$ (Blum et al.\ 1996), using 
either the Indebetouw et al.\ (2005) or Flaherty et al.\ (2007) extinction laws, we should only 
detect sources with absolute magnitudes $\lesssim 1$. Variations in extinction across the region 
may alter the apparent magnitudes of some IR sources, but the main limitation of the IRAC survey
is due to confusion, not extinction nor sensitivity. At 8 $\micron$ and to a lesser degree at 5.8 
$\micron$, a large instrumental beam, and confusion from the diffuse ISM are additional limitations
which reduce the number of sources detected compared to 3.6 and 4.5 $\micron$.
While the extinction is not the dominant factor in the detection of the IR sources, it is interesting 
to note that extinction appears to have a strong effect on the distribution of the X--ray sources.
In Figures \ref{fig_ch1x}a and \ref{fig_ch4x}a, there is an evident correlation between 
a relatively low X--ray source density and an IR dark cloud (IRDC) to the southeast of 
the Galactic Center.

\section{Conclusions}

We have analyzed the possible correlations between the largest number of candidate 
sources to date: 2357 X--ray sources (of which 1809 are hard X--ray sources most 
likely located at the Galactic Center) and $\sim$20,000 {\it Spitzer}/IRAC infrared point sources.
Source confusion limits our correlations
to only bright infrared sources with absolute magnitudes $\lesssim 1$ if located near the Galactic Center.
The lack of any significant correlation between hard X--ray sources and 
3.6 -- 8 $\micron$ infrared point sources suggests that there is no unique population of sources that 
are bright at both X--ray and 3.6 -- 8 $\micron$ wavelengths. Based on this study, we can set the upper limit 
on the fraction of all hard X--ray sources that can be bright at both X--ray and 3.6 -- 8 $\micron$ wavelengths 
to be $<1.7\%$ (3$\sigma$).

\acknowledgements
This work is based on observations made with the Spitzer Space Telescope, which is
operated by the Jet Propulsion Laboratory, California Institute of Technology under
a contract with NASA. Support for this work was provided by NASA.
KS thanks the NASA Faculty Fellowship Program for financial support
and the hospitality of JPL's Long Wavelength Center and the Spitzer Science Center. 
We thank the referee for substantial comments and suggestions.

\clearpage

\begin{deluxetable}{lccccccc}
\tabletypesize{\scriptsize}
\tablewidth{0pt}
\tablecaption{All X-ray Sources (2357)}
\tablehead{
\colhead{Wavelength ($\micron$)} &
\colhead{$N$} & 
\colhead{$\sigma_N$} & 
\colhead{$M$} & 
\colhead{$\sigma_M$} & 
\colhead{$N-M$} & 
\colhead{$\sigma_{N-M}$} & 
\colhead{$(N-M)/\sigma_{N-M}$} 
}
\startdata
\cutinhead{Positional agreement $< 0.5''$}
3.6 &     51 &  7.1 &      30.5 &  2.0 &      20.5 &  7.4 &  2.8\\
4.5 &     46 &  6.8 &      27.8 &  1.9 &      18.2 &  7.0 &  2.6\\
5.8 &     35 &  5.9 &      19.9 &  1.6 &      15.1 &  6.1 &  2.5\\
8   &     22 &  4.7 &      10.1 &  1.1 &      11.9 &  4.8 &  2.5\\

\cutinhead{Positional agreement $< 1.0''$}
3.6 &    156 & 12.5 &     120.4 &  3.9 &      35.6 & 13.1 &  2.7\\
4.5 &    137 & 11.7 &     108.2 &  3.7 &      28.8 & 12.3 &  2.3\\
5.8 &    104 & 10.2 &      77.4 &  3.1 &      26.6 & 10.7 &  2.5\\
8   &     64 &  8.0 &      42.0 &  2.3 &      22.0 &  8.3 &  2.6\\

\cutinhead{Positional agreement $< 2.0''$}
3.6 &    488 & 22.1 &     458.4 &  7.6 &      29.6 & 23.4 &  1.3\\
4.5 &    424 & 20.6 &     414.5 &  7.2 &       9.5 & 21.8 &  0.4\\
5.8 &    320 & 17.9 &     297.9 &  6.1 &      22.1 & 18.9 &  1.2\\
8   &    179 & 13.4 &     170.9 &  4.6 &       8.1 & 14.2 &  0.6\\
\enddata
\end{deluxetable}

\begin{deluxetable}{lccccccc}
\tabletypesize{\scriptsize}
\tablewidth{0pt}
\tablecaption{All X-ray Sources with Uncertainty $<\ 0\farcs8$ (1645)}
\tablehead{
\colhead{Wavelength ($\micron$)} &
\colhead{$N$} & 
\colhead{$\sigma_N$} & 
\colhead{$M$} & 
\colhead{$\sigma_M$} & 
\colhead{$N-M$} & 
\colhead{$\sigma_{N-M}$} & 
\colhead{$(N-M)/\sigma_{N-M}$} 
}
\startdata
\cutinhead{Positional agreement $< 0.5''$}
3.6  &    29 &  5.4  &     20.5 &  1.6  &      8.5 &  5.6 &  1.5\\
4.5  &    27 &  5.2  &     18.2 &  1.5  &      8.8 &  5.4 &  1.6\\
5.8  &    19 &  4.4  &     13.0 &  1.3  &      6.0 &  4.5 &  1.3\\
8.0  &    11 &  3.3  &      6.2 &  0.9  &      4.8 &  3.4 &  1.4\\
\cutinhead{Positional agreement $< 1.0''$}
3.6  &    97 &  9.8  &     86.0 &  3.3  &     11.0 & 10.4 &  1.1\\
4.5  &    86 &  9.3  &     75.8 &  3.1  &     10.2 &  9.8 &  1.0\\
5.8  &    66 &  8.1  &     55.5 &  2.6  &     10.5 &  8.5 &  1.2\\
8.0  &    37 &  6.1  &     29.4 &  1.9  &      7.6 &  6.4 &  1.2\\
\cutinhead{Positional agreement $< 2.0''$}
3.6  &   328 & 18.1  &    324.9 &  6.4  &      3.1 & 19.2 &  0.2\\
4.5  &   272 & 16.5  &    290.9 &  6.0  &    -18.9 & 17.6 & -1.1\\
5.8  &   213 & 14.6  &    210.4 &  5.1  &      2.6 & 15.5 &  0.2\\
8.0  &   121 & 11.0  &    118.6 &  3.9  &      2.4 & 11.7 &  0.2\\
\enddata
\end{deluxetable}

\begin{deluxetable}{lccccccc}
\tabletypesize{\scriptsize}
\tablewidth{0pt}
\tablecaption{Soft X-ray Sources (548)}
\tablehead{
\colhead{Wavelength ($\micron$)} &
\colhead{$N$} & 
\colhead{$\sigma_N$} & 
\colhead{$M$} & 
\colhead{$\sigma_M$} & 
\colhead{$N-M$} & 
\colhead{$\sigma_{N-M}$} & 
\colhead{$(N-M)/\sigma_{N-M}$} 
}
\startdata
\cutinhead{Positional agreement $< 0.5''$}
3.6 & 28 & 5.3 & 5.8 & 0.8 & 22.2 & 5.4 & 4.2\\
4.8 & 23 & 4.8 & 5.1 & 0.8 & 17.9 & 4.9 & 3.7\\
5.8 & 17 & 4.1 & 3.5 & 0.7 & 13.5 & 4.2 & 3.2\\
8   & 8 & 2.8 & 1.5 & 0.4 & 6.5 & 2.9 & 2.3\\

\cutinhead{Positional agreement $< 1.0''$}
3.6 & 64 & 8.0 & 26.4 & 1.8 & 37.6 & 8.2 & 4.6\\
4.5 & 55 & 7.4 & 24.1 & 1.7 & 30.9 & 7.6 & 4.1\\
5.8 & 37 & 6.1 & 15.9 & 1.4 & 21.1 & 6.2 & 3.4\\
8   & 20 & 4.5 & 7.2 & 1.0 & 12.8 & 4.6 & 2.8\\

\cutinhead{Positional agreement $< 2.0''$}
3.6 & 126 & 11.2 & 103.0 & 3.6 & 23.0 & 11.8 & 2.0\\
4.5 & 109 & 10.4 & 95.1 & 3.4 & 13.9 & 11.0 & 1.3\\
5.8 & 78 & 8.8 & 69.2 & 2.9 & 8.8 & 9.3 & 0.9\\
8   & 42 & 6.5 & 37.4 & 2.2 & 4.6 & 6.8 & 0.7\\
\enddata
\end{deluxetable}

\begin{deluxetable}{lccccccc}
\tabletypesize{\scriptsize}
\tablewidth{0pt}
\tablecaption{Soft X-ray Sources with Uncertainty $<\ 0\farcs8$ (303)}
\tablehead{
\colhead{Wavelength ($\micron$)} &
\colhead{$N$} & 
\colhead{$\sigma_N$} & 
\colhead{$M$} & 
\colhead{$\sigma_M$} & 
\colhead{$N-M$} & 
\colhead{$\sigma_{N-M}$} & 
\colhead{$(N-M)/\sigma_{N-M}$} 
}
\startdata
\cutinhead{Positional agreement $< 0.5''$}
3.6  &    16  & 4.0   &     3.2 &  0.6    &   12.8 &  4.1 &  3.1\\
4.5  &    14  & 3.7   &     2.9 &  0.6    &   11.1 &  3.8 &  2.9\\
5.8  &     9  & 3.0   &     1.6 &  0.5    &    7.4 &  3.0 &  2.4\\
8.0  &     3  & 1.7   &     0.9 &  0.3    &    2.1 &  1.8 &  1.2\\
\cutinhead{Positional agreement $< 1.0''$}
3.6  &    34  & 5.8   &    15.4 &  1.4    &   18.6 &  6.0 &  3.1\\
4.5  &    30  & 5.5   &    13.9 &  1.3    &   16.1 &  5.6 &  2.9\\
5.8  &    19  & 4.4   &     8.6 &  1.0    &   10.4 &  4.5 &  2.3\\
8.0  &     9  & 3.0   &     4.2 &  0.7    &    4.8 &  3.1 &  1.5\\
\cutinhead{Positional agreement $< 2.0''$}
3.6  &    68  & 8.2   &    58.4 &  2.7    &    9.6 &  8.7 &  1.1\\
4.5  &    57  & 7.5   &    54.0 &  2.6    &    3.0 &  8.0 &  0.4\\
5.8  &    40  & 6.3   &    37.4 &  2.2    &    2.6 &  6.7 &  0.4\\
8.0  &    21  & 4.6   &    21.0 &  1.6    &    0.0 &  4.9 &  0.0\\
\enddata
\end{deluxetable}

\begin{deluxetable}{lccccccc}
\tabletypesize{\scriptsize}
\tablewidth{0pt}
\tablecaption{Hard X-ray Sources (1809)}
\tablehead{
\colhead{Wavelength ($\micron$)} &
\colhead{$N$} & 
\colhead{$\sigma_N$} & 
\colhead{$M$} & 
\colhead{$\sigma_M$} & 
\colhead{$N-M$} & 
\colhead{$\sigma_{N-M}$} & 
\colhead{$(N-M)/\sigma_{N-M}$} 
}
\startdata
\cutinhead{Positional agreement $< 0.5''$}
3.6 & 23 & 4.8 & 24.8 & 1.8 & -1.8 & 5.1 & -0.3\\
4.5 & 23 & 4.8 & 22.6 & 1.7 & 0.4 & 5.1 & 0.1\\
5.8 & 18 & 4.2 & 16.4 & 1.4 & 1.6 & 4.5 & 0.4\\
8   & 14 & 3.7 & 8.6 & 1.0 & 5.4 & 3.9 & 1.4\\

\cutinhead{Positional agreement $< 1.0''$}
3.6 & 92 & 9.6 & 94.0 & 3.4 & -2.0 & 10.2 & -0.2\\
4.5 & 82 & 9.1 & 84.1 & 3.2 & -2.1 & 9.6 & -0.2\\
5.8 & 67 & 8.2 & 61.5 & 2.8 & 5.5 & 8.6 & 0.6\\
8   & 44 & 6.6 & 34.8 & 2.1 & 9.2 & 7.0 & 1.3\\

\cutinhead{Positional agreement $< 2.0''$}
3.6 & 362 & 19.0 & 355.4 & 6.7 & 6.6 & 20.2 & 0.3\\
4.5 & 315 & 17.7 & 319.4 & 6.3 & -4.4 & 18.8 & -0.2\\
5.8 & 242 & 15.6 & 228.6 & 5.3 & 13.4 & 16.4 & 0.8\\
8   & 137 & 11.7 & 133.5 & 4.1 & 3.5 & 12.4 & 0.3\\
\enddata
\end{deluxetable}

\begin{deluxetable}{lccccccc}
\tabletypesize{\scriptsize}
\tablewidth{0pt}
\tablecaption{Hard X-ray Sources with Uncertainty $<\ 0\farcs8$ (1342)}
\tablehead{
\colhead{Wavelength ($\micron$)} &
\colhead{$N$} & 
\colhead{$\sigma_N$} & 
\colhead{$M$} & 
\colhead{$\sigma_M$} & 
\colhead{$N-M$} & 
\colhead{$\sigma_{N-M}$} & 
\colhead{$(N-M)/\sigma_{N-M}$} 
}
\startdata
\cutinhead{Positional agreement $< 0.5''$}
3.6 &     13 &  3.6  &     17.2 &  1.5  &     -4.2 &  3.9 & -1.1\\
4.5 &     13 &  3.6  &     15.4 &  1.4  &     -2.4 &  3.9 & -0.6\\
5.8 &     10 &  3.2  &     11.4 &  1.2  &     -1.4 &  3.4 & -0.4\\
8.0 &      8 &  2.8  &      5.4 &  0.8  &      2.6 &  2.9 &  0.9\\
\cutinhead{Positional agreement $< 1.0''$}
3.6 &     63 &  7.9  &     70.6 &  3.0  &     -7.6 &  8.5 & -0.9\\
4.5 &     56 &  7.5  &     61.9 &  2.8  &     -5.9 &  8.0 & -0.7\\
5.8 &     47 &  6.9  &     46.9 &  2.4  &      0.1 &  7.3 &  0.0\\
8.0 &     28 &  5.3  &     25.1 &  1.8  &      2.9 &  5.6 &  0.5\\
\cutinhead{Positional agreement $< 2.0''$}
3.6 &    260 & 16.1  &    266.5 &  5.8  &     -6.5 & 17.1 & -0.4\\
4.5 &    215 & 14.7  &    236.9 &  5.4  &    -21.9 & 15.6 & -1.4\\
5.8 &    173 & 13.2  &    173.0 &  4.7  &      0.0 & 14.0 &  0.0\\
8.0 &    100 & 10.0  &     97.6 &  3.5  &      2.4 & 10.6 &  0.2\\
\enddata
\end{deluxetable}


\begin{deluxetable}{lllccccclllcl}
\rotate
\tabletypesize{\scriptsize}
\tablewidth{0pt}
\tablecaption{IR and X--Ray Sources Coinciding within $1''$ (No Physical Association Implied)}
\tablehead{
\multicolumn{7}{c}{IR Sources (Ram\'irez at al. 2008)} & & \multicolumn{4}{c}{X--Ray Sources (Muno et al.\ 2003)}\\
\cline{1-7}\cline{9-12}
\colhead{Name} &
\colhead{R.A. (deg)} & 
\colhead{Dec. (deg)} & 
\colhead{[3.6]} & 
\colhead{[4.5]} & 
\colhead{[5.8]} & 
\colhead{[8.0]} & 
\colhead{}&
\colhead{Name} &
\colhead{R.A. (deg) } &
\colhead{Dec. (deg)} &
\colhead{Spectrum\tablenotemark{a}} &
\colhead{$\Delta$ ($''$)\tablenotemark{b}} 
}
\startdata
SSTGC 0469964 & 266.33191 & -29.02930 &     11.54 &     11.59 &   \nodata &   \nodata & &  CXOGCJ174519.7-290146 &    266.33212 & -29.02947 & soft &      0.91 \\
SSTGC 0470990 & 266.33348 & -29.04629 &     11.15 &   \nodata &   \nodata &   \nodata & &  CXOGCJ174520.0-290246 &    266.33374 & -29.04620 & hard &      0.87 \\
SSTGC 0473500 & 266.33735 & -29.03927 &      8.91 &      8.17 &   \nodata &   \nodata & &  CXOGCJ174521.0-290221 &    266.33755 & -29.03930 & hard &      0.64 \\
SSTGC 0476142 & 266.34154 & -28.99323 &      9.85 &      9.68 &   \nodata &   \nodata & &  CXOGCJ174521.9-285936 &    266.34134 & -28.99344 & hard &      0.99 \\
SSTGC 0476234 & 266.34169 & -29.00903 &      9.42 &      8.95 &      8.30 &   \nodata & &  CXOGCJ174521.9-290032 &    266.34159 & -29.00908 & soft &      0.36 \\
SSTGC 0477415 & 266.34345 & -29.01875 &      9.28 &      9.21 &      8.77 &      8.55 & &  CXOGCJ174522.4-290107 &    266.34369 & -29.01881 & hard &      0.79 \\
SSTGC 0479734 & 266.34707 & -28.95763 &      9.43 &      9.33 &      8.90 &   \nodata & &  CXOGCJ174523.2-285727 &    266.34682 & -28.95750 & soft &      0.92 \\
SSTGC 0479445 & 266.34664 & -29.01752 &     10.12 &      9.34 &      9.28 &      9.41 & &  CXOGCJ174523.2-290103 &    266.34684 & -29.01764 & hard &      0.77 \\
SSTGC 0481996 & 266.35058 & -28.97954 &      9.93 &      9.51 &      9.21 &      9.32 & &  CXOGCJ174524.1-285845 &    266.35064 & -28.97932 & soft &      0.82 \\
SSTGC 0482409 & 266.35126 & -29.03542 &     10.02 &      9.90 &      9.44 &   \nodata & &  CXOGCJ174524.3-290208 &    266.35145 & -29.03561 & soft &      0.91 \\
SSTGC 0483850 & 266.35344 & -29.00869 &     10.44 &   \nodata &   \nodata &   \nodata & &  CXOGCJ174524.7-290031 &    266.35326 & -29.00888 & hard &      0.89 \\
SSTGC 0484205 & 266.35400 & -29.04215 &      9.62 &      9.57 &      9.28 &   \nodata & &  CXOGCJ174525.0-290232 &    266.35419 & -29.04226 & hard &      0.71 \\
SSTGC 0485202 & 266.35557 & -29.02035 &     11.64 &     11.09 &   \nodata &   \nodata & &  CXOGCJ174525.2-290113 &    266.35539 & -29.02049 & hard &      0.75 \\
SSTGC 0487698 & 266.35948 & -28.99733 &      8.92 &      8.10 &      7.59 &      7.51 & &  CXOGCJ174526.3-285949 &    266.35961 & -28.99721 & hard &      0.60 \\
SSTGC 0492659 & 266.36710 & -28.99540 &   \nodata &   \nodata &   \nodata &      9.23 & &  CXOGCJ174528.0-285943 &    266.36702 & -28.99547 & hard &      0.36 \\
SSTGC 0492380 & 266.36669 & -29.00646 &     10.18 &      9.96 &   \nodata &   \nodata & &  CXOGCJ174528.0-290023 &    266.36680 & -29.00644 & soft &      0.36 \\
SSTGC 0493525 & 266.36850 & -29.05255 &     11.22 &     10.93 &   \nodata &   \nodata & &  CXOGCJ174528.4-290308 &    266.36841 & -29.05230 & hard &      0.93 \\
SSTGC 0494379 & 266.36984 & -28.99134 &     10.28 &   \nodata &      9.88 &   \nodata & &  CXOGCJ174528.7-285928 &    266.36960 & -28.99120 & hard &      0.91 \\
SSTGC 0494207 & 266.36958 & -29.01202 &      9.74 &      9.65 &      9.07 &      8.93 & &  CXOGCJ174528.7-290042 &    266.36979 & -29.01183 & hard &      0.96 \\
SSTGC 0494633 & 266.37021 & -28.95736 &     10.68 &     10.21 &     10.10 &   \nodata & &  CXOGCJ174528.8-285726 &    266.37030 & -28.95749 & soft &      0.55 \\
SSTGC 0495190 & 266.37111 & -29.06847 &      6.41 &      5.81 &      5.19 &      5.36 & &  CXOGCJ174529.0-290406 &    266.37109 & -29.06846 & hard &      0.08 \\
SSTGC 0496104 & 266.37251 & -29.00714 &     11.67 &     11.38 &   \nodata &   \nodata & &  CXOGCJ174529.4-290025 &    266.37254 & -29.00699 & hard &      0.56 \\
SSTGC 0496570 & 266.37324 & -29.03733 &      9.47 &      9.42 &   \nodata &   \nodata & &  CXOGCJ174529.5-290215 &    266.37326 & -29.03754 & soft &      0.75 \\
SSTGC 0496855 & 266.37365 & -28.95047 &     11.79 &   \nodata &   \nodata &   \nodata & &  CXOGCJ174529.6-285701 &    266.37370 & -28.95046 & hard &      0.16 \\
SSTGC 0496842 & 266.37363 & -28.97793 &     10.30 &      9.69 &      9.50 &   \nodata & &  CXOGCJ174529.6-285840 &    266.37370 & -28.97790 & hard &      0.25 \\
SSTGC 0496808 & 266.37358 & -29.00602 &     10.78 &     10.42 &   \nodata &   \nodata & &  CXOGCJ174529.6-290021 &    266.37375 & -29.00607 & hard &      0.57 \\
SSTGC 0496815 & 266.37359 & -29.04089 &      9.32 &      9.20 &   \nodata &   \nodata & &  CXOGCJ174529.6-290227 &    266.37363 & -29.04084 & hard &      0.23 \\
SSTGC 0498139 & 266.37558 & -29.04694 &      9.15 &      8.53 &      8.12 &      8.32 & &  CXOGCJ174530.0-290248 &    266.37530 & -29.04681 & hard &      1.00 \\
SSTGC 0498461 & 266.37611 & -29.06168 &     10.29 &     10.24 &     10.02 &   \nodata & &  CXOGCJ174530.3-290341 &    266.37628 & -29.06157 & hard &      0.67 \\
SSTGC 0500061 & 266.37851 & -29.02794 &      7.03 &      5.84 &      5.04 &      4.79 & &  CXOGCJ174530.8-290139 &    266.37864 & -29.02777 & hard &      0.73 \\
SSTGC 0502935 & 266.38299 & -29.04974 &      7.01 &      6.92 &      6.51 &      6.44 & &  CXOGCJ174531.9-290258 &    266.38306 & -29.04952 & hard &      0.83 \\
SSTGC 0503729 & 266.38422 & -29.02490 &   \nodata &      9.80 &   \nodata &   \nodata & &  CXOGCJ174532.1-290130 &    266.38416 & -29.02509 & hard &      0.72 \\
SSTGC 0503925 & 266.38452 & -29.05826 &   \nodata &     10.48 &   \nodata &   \nodata & &  CXOGCJ174532.2-290329 &    266.38439 & -29.05829 & hard &      0.43 \\
SSTGC 0504458 & 266.38536 & -28.98557 &      9.86 &      9.96 &      9.64 &   \nodata & &  CXOGCJ174532.5-285908 &    266.38566 & -28.98556 & hard &      0.94 \\
SSTGC 0505766 & 266.38734 & -29.07664 &     10.91 &     10.56 &   \nodata &   \nodata & &  CXOGCJ174532.8-290436 &    266.38703 & -29.07670 & hard &      1.00 \\
SSTGC 0507145 & 266.38945 & -28.94431 &     10.81 &     10.43 &      9.89 &   \nodata & &  CXOGCJ174533.4-285638 &    266.38941 & -28.94415 & soft &      0.58 \\
SSTGC 0508028 & 266.39072 & -28.95800 &     11.45 &     10.97 &   \nodata &   \nodata & &  CXOGCJ174533.7-285728 &    266.39077 & -28.95803 & soft &      0.19 \\
SSTGC 0507814 & 266.39043 & -29.07579 &      9.62 &      8.73 &      8.37 &   \nodata & &  CXOGCJ174533.7-290432 &    266.39053 & -29.07578 & hard &      0.32 \\
SSTGC 0508562 & 266.39155 & -28.99906 &      9.42 &   \nodata &   \nodata &   \nodata & &  CXOGCJ174534.0-285956 &    266.39182 & -28.99901 & hard &      0.87 \\
SSTGC 0510334 & 266.39416 & -29.04338 &      7.98 &      7.73 &      7.17 &      6.47 & &  CXOGCJ174534.5-290236 &    266.39409 & -29.04357 & hard &      0.73 \\
SSTGC 0510970 & 266.39515 & -29.07890 &      7.48 &      7.39 &      6.80 &      6.98 & &  CXOGCJ174534.8-290444 &    266.39535 & -29.07906 & hard &      0.84 \\
SSTGC 0511940 & 266.39659 & -28.97869 &   \nodata &   \nodata &      8.27 &   \nodata & &  CXOGCJ174535.1-285843 &    266.39633 & -28.97877 & hard &      0.87 \\
SSTGC 0511999 & 266.39668 & -29.01359 &      8.57 &      8.45 &      7.86 &      7.22 & &  CXOGCJ174535.2-290048 &    266.39689 & -29.01352 & hard &      0.71 \\
SSTGC 0512567 & 266.39755 & -28.93489 &      9.50 &      9.57 &      9.27 &   \nodata & &  CXOGCJ174535.4-285605 &    266.39763 & -28.93492 & hard &      0.28 \\
SSTGC 0514496 & 266.40052 & -28.94409 &      8.82 &      8.26 &      7.97 &      7.82 & &  CXOGCJ174536.1-285638 &    266.40059 & -28.94407 & soft &      0.23 \\
SSTGC 0514528 & 266.40057 & -28.96362 &     10.20 &     10.36 &      9.89 &   \nodata & &  CXOGCJ174536.1-285748 &    266.40067 & -28.96347 & hard &      0.62 \\
SSTGC 0514886 & 266.40111 & -28.97721 &   \nodata &   \nodata &   \nodata &      8.17 & &  CXOGCJ174536.3-285837 &    266.40125 & -28.97712 & hard &      0.55 \\
SSTGC 0515077 & 266.40139 & -29.02929 &     10.73 &     10.41 &   \nodata &   \nodata & &  CXOGCJ174536.3-290145 &    266.40150 & -29.02919 & soft &      0.51 \\
SSTGC 0515409 & 266.40189 & -29.02083 &      9.30 &   \nodata &   \nodata &   \nodata & &  CXOGCJ174536.4-290114 &    266.40181 & -29.02067 & hard &      0.64 \\
SSTGC 0517465 & 266.40502 & -28.98550 &      8.15 &      8.16 &      7.79 &   \nodata & &  CXOGCJ174537.2-285906 &    266.40502 & -28.98526 & hard &      0.87 \\
SSTGC 0518345 & 266.40641 & -29.03129 &     10.04 &      9.92 &      9.80 &   \nodata & &  CXOGCJ174537.5-290153 &    266.40653 & -29.03142 & hard &      0.62 \\
SSTGC 0518659 & 266.40684 & -29.00667 &      9.45 &   \nodata &   \nodata &   \nodata & &  CXOGCJ174537.6-290023 &    266.40683 & -29.00657 & hard &      0.36 \\
SSTGC 0519255 & 266.40776 & -29.02434 &      9.93 &      9.76 &      8.65 &      7.41 & &  CXOGCJ174537.8-290127 &    266.40757 & -29.02426 & soft &      0.66 \\
SSTGC 0519545 & 266.40824 & -29.02626 &      8.25 &      8.13 &   \nodata &   \nodata & &  CXOGCJ174537.9-290134 &    266.40831 & -29.02622 & soft &      0.26 \\
SSTGC 0520087 & 266.40907 & -28.98803 &   \nodata &   \nodata &      8.26 &   \nodata & &  CXOGCJ174538.1-285916 &    266.40878 & -28.98805 & hard &      0.92 \\
SSTGC 0520903 & 266.41025 & -28.96968 &      8.50 &      8.45 &      7.91 &      7.69 & &  CXOGCJ174538.4-285810 &    266.41031 & -28.96958 & hard &      0.40 \\
SSTGC 0521476 & 266.41115 & -29.01893 &   \nodata &      8.51 &   \nodata &   \nodata & &  CXOGCJ174538.6-290107 &    266.41101 & -29.01889 & hard &      0.47 \\
SSTGC 0522739 & 266.41312 & -28.96061 &      9.19 &      9.22 &      8.79 &      8.46 & &  CXOGCJ174539.1-285738 &    266.41308 & -28.96081 & hard &      0.73 \\
SSTGC 0522856 & 266.41327 & -29.04730 &     11.06 &      8.88 &      8.23 &      8.20 & &  CXOGCJ174539.2-290250 &    266.41356 & -29.04738 & hard &      0.95 \\
SSTGC 0523283 & 266.41392 & -29.00467 &      8.12 &   \nodata &   \nodata &   \nodata & &  CXOGCJ174539.3-290016 &    266.41397 & -29.00467 & soft &      0.16 \\
SSTGC 0523729 & 266.41463 & -29.02514 &      9.49 &      8.94 &      8.81 &   \nodata & &  CXOGCJ174539.5-290129 &    266.41465 & -29.02499 & hard &      0.54 \\
SSTGC 0524504 & 266.41584 & -29.00847 &      5.63 &   \nodata &   \nodata &   \nodata & &  CXOGCJ174539.7-290029 &    266.41567 & -29.00827 & soft &      0.91 \\
SSTGC 0525460 & 266.41726 & -28.99969 &      6.18 &      4.59 &      3.16 &      1.69 & &  CXOGCJ174540.1-285957 &    266.41718 & -28.99944 & hard &      0.92 \\
SSTGC 0525510 & 266.41735 & -29.01540 &      8.21 &   \nodata &      6.98 &   \nodata & &  CXOGCJ174540.1-290055 &    266.41733 & -29.01546 & soft &      0.23 \\
SSTGC 0525714 & 266.41765 & -28.98358 &      8.52 &      8.46 &      7.72 &      6.99 & &  CXOGCJ174540.2-285900 &    266.41753 & -28.98360 & hard &      0.38 \\
SSTGC 0526867 & 266.41944 & -28.94462 &   \nodata &   \nodata &   \nodata &      8.58 & &  CXOGCJ174540.6-285640 &    266.41950 & -28.94445 & soft &      0.65 \\
SSTGC 0527316 & 266.42016 & -29.03007 &   \nodata &   \nodata &      8.12 &   \nodata & &  CXOGCJ174540.8-290149 &    266.42011 & -29.03029 & hard &      0.80 \\
SSTGC 0527347 & 266.42020 & -29.05794 &     11.12 &   \nodata &   \nodata &   \nodata & &  CXOGCJ174540.8-290328 &    266.42037 & -29.05804 & soft &      0.65 \\
SSTGC 0527844 & 266.42100 & -29.00472 &      8.56 &      7.16 &   \nodata &   \nodata & &  CXOGCJ174541.0-290017 &    266.42095 & -29.00489 & hard &      0.62 \\
SSTGC 0529789 & 266.42400 & -29.00790 &      9.01 &   \nodata &   \nodata &   \nodata & &  CXOGCJ174541.7-290027 &    266.42377 & -29.00774 & hard &      0.93 \\
SSTGC 0530322 & 266.42482 & -28.99881 &   \nodata &   \nodata &      8.47 &   \nodata & &  CXOGCJ174541.9-285955 &    266.42473 & -28.99865 & hard &      0.65 \\
SSTGC 0531615 & 266.42680 & -28.96723 &   \nodata &   \nodata &   \nodata &      9.28 & &  CXOGCJ174542.4-285802 &    266.42678 & -28.96724 & hard &      0.07 \\
SSTGC 0532179 & 266.42763 & -28.95284 &     11.30 &     10.80 &   \nodata &   \nodata & &  CXOGCJ174542.6-285709 &    266.42768 & -28.95257 & hard &      0.99 \\
SSTGC 0532570 & 266.42818 & -28.97533 &      9.54 &      9.29 &      8.87 &   \nodata & &  CXOGCJ174542.7-285831 &    266.42816 & -28.97552 & hard &      0.68 \\
SSTGC 0533119 & 266.42903 & -29.07535 &   \nodata &     10.74 &   \nodata &   \nodata & &  CXOGCJ174542.9-290431 &    266.42890 & -29.07535 & soft &      0.41 \\
SSTGC 0533640 & 266.42983 & -29.01391 &      7.39 &      6.73 &      6.11 &      5.85 & &  CXOGCJ174543.1-290049 &    266.42961 & -29.01382 & soft &      0.76 \\
SSTGC 0533886 & 266.43016 & -28.98839 &     10.20 &     10.18 &      9.59 &   \nodata & &  CXOGCJ174543.2-285917 &    266.43035 & -28.98822 & hard &      0.85 \\
SSTGC 0534588 & 266.43123 & -28.98360 &      9.10 &   \nodata &      9.08 &   \nodata & &  CXOGCJ174543.4-285900 &    266.43102 & -28.98341 & hard &      0.96 \\
SSTGC 0534566 & 266.43120 & -29.06347 &      8.47 &      8.36 &      7.85 &      7.97 & &  CXOGCJ174543.4-290347 &    266.43106 & -29.06327 & hard &      0.83 \\
SSTGC 0535312 & 266.43230 & -29.02660 &     11.89 &     10.66 &   \nodata &   \nodata & &  CXOGCJ174543.7-290136 &    266.43234 & -29.02668 & hard &      0.32 \\
SSTGC 0535628 & 266.43280 & -29.04581 &      9.84 &      9.81 &      9.31 &   \nodata & &  CXOGCJ174543.9-290245 &    266.43309 & -29.04590 & hard &      0.96 \\
SSTGC 0535765 & 266.43300 & -29.08237 &      9.46 &      9.48 &      9.43 &      9.48 & &  CXOGCJ174543.9-290456 &    266.43305 & -29.08238 & soft &      0.16 \\
SSTGC 0536043 & 266.43341 & -29.07468 &     10.24 &      9.67 &      9.44 &      9.04 & &  CXOGCJ174544.0-290428 &    266.43363 & -29.07460 & hard &      0.75 \\
SSTGC 0537314 & 266.43533 & -28.97495 &      9.36 &      8.80 &      8.24 &      7.83 & &  CXOGCJ174544.4-285829 &    266.43531 & -28.97486 & hard &      0.32 \\
SSTGC 0537395 & 266.43544 & -28.97053 &   \nodata &   \nodata &      6.76 &   \nodata & &  CXOGCJ174544.5-285813 &    266.43557 & -28.97044 & hard &      0.52 \\
SSTGC 0537801 & 266.43609 & -29.06095 &     12.55 &   \nodata &   \nodata &   \nodata & &  CXOGCJ174544.7-290339 &    266.43631 & -29.06109 & hard &      0.87 \\
SSTGC 0539467 & 266.43859 & -28.97468 &     10.76 &     10.45 &   \nodata &   \nodata & &  CXOGCJ174545.2-285828 &    266.43870 & -28.97466 & soft &      0.36 \\
SSTGC 0539388 & 266.43847 & -29.04029 &     12.23 &     12.24 &   \nodata &   \nodata & &  CXOGCJ174545.2-290224 &    266.43840 & -29.04009 & soft &      0.75 \\
SSTGC 0543526 & 266.44459 & -29.05783 &     10.56 &   \nodata &   \nodata &   \nodata & &  CXOGCJ174546.6-290328 &    266.44449 & -29.05786 & soft &      0.33 \\
SSTGC 0543690 & 266.44482 & -29.03528 &   \nodata &   \nodata &   \nodata &      8.35 & &  CXOGCJ174546.7-290207 &    266.44473 & -29.03552 & hard &      0.92 \\
SSTGC 0544019 & 266.44530 & -29.04789 &      9.47 &      8.58 &      7.83 &   \nodata & &  CXOGCJ174546.8-290252 &    266.44521 & -29.04789 & soft &      0.28 \\
SSTGC 0546444 & 266.44903 & -28.94019 &      7.82 &      7.48 &      6.85 &      6.78 & &  CXOGCJ174547.7-285624 &    266.44899 & -28.94008 & hard &      0.41 \\
SSTGC 0547195 & 266.45015 & -29.05055 &      7.54 &      7.26 &      6.80 &      6.89 & &  CXOGCJ174548.0-290301 &    266.45010 & -29.05040 & hard &      0.55 \\
SSTGC 0548119 & 266.45158 & -28.99872 &   \nodata &   \nodata &      9.94 &   \nodata & &  CXOGCJ174548.4-285954 &    266.45179 & -28.99852 & hard &      0.97 \\
SSTGC 0548472 & 266.45212 & -28.96311 &     10.88 &     10.62 &   \nodata &   \nodata & &  CXOGCJ174548.5-285747 &    266.45240 & -28.96309 & hard &      0.88 \\
SSTGC 0548665 & 266.45244 & -29.04621 &     11.93 &   \nodata &   \nodata &   \nodata & &  CXOGCJ174548.5-290246 &    266.45248 & -29.04616 & soft &      0.23 \\
SSTGC 0550312 & 266.45497 & -29.02723 &     10.71 &     10.72 &     10.67 &   \nodata & &  CXOGCJ174549.1-290137 &    266.45492 & -29.02721 & soft &      0.18 \\
SSTGC 0550210 & 266.45481 & -29.08411 &     10.33 &      9.94 &      9.58 &      9.10 & &  CXOGCJ174549.1-290502 &    266.45467 & -29.08396 & soft &      0.71 \\
SSTGC 0550863 & 266.45578 & -28.95373 &      8.27 &      8.02 &      7.54 &      7.52 & &  CXOGCJ174549.4-285712 &    266.45597 & -28.95356 & hard &      0.87 \\
SSTGC 0553475 & 266.45968 & -29.00568 &     12.25 &     11.87 &   \nodata &   \nodata & &  CXOGCJ174550.2-290020 &    266.45948 & -29.00566 & hard &      0.63 \\
SSTGC 0554271 & 266.46090 & -28.98878 &      9.63 &      9.00 &      8.63 &      7.90 & &  CXOGCJ174550.6-285919 &    266.46096 & -28.98886 & soft &      0.36 \\
SSTGC 0554555 & 266.46134 & -29.07635 &     12.52 &     11.67 &   \nodata &   \nodata & &  CXOGCJ174550.7-290434 &    266.46159 & -29.07628 & soft &      0.83 \\
SSTGC 0558398 & 266.46721 & -28.96776 &      9.82 &      9.85 &      9.51 &   \nodata & &  CXOGCJ174552.1-285804 &    266.46722 & -28.96790 & soft &      0.50 \\
SSTGC 0559457 & 266.46887 & -28.99130 &   \nodata &   \nodata &      7.42 &   \nodata & &  CXOGCJ174552.5-285928 &    266.46878 & -28.99131 & hard &      0.29 \\
SSTGC 0560119 & 266.46985 & -29.03013 &     10.41 &   \nodata &   \nodata &   \nodata & &  CXOGCJ174552.7-290148 &    266.46965 & -29.03021 & hard &      0.69 \\
SSTGC 0564995 & 266.47723 & -29.04817 &     11.86 &     11.62 &   \nodata &   \nodata & &  CXOGCJ174554.5-290252 &    266.47709 & -29.04805 & soft &      0.62 \\
SSTGC 0570573 & 266.48573 & -28.96011 &     10.59 &     10.53 &     11.08 &   \nodata & &  CXOGCJ174556.5-285736 &    266.48560 & -28.96019 & soft &      0.50 \\
SSTGC 0572280 & 266.48837 & -29.05031 &      9.12 &      9.07 &      8.67 &   \nodata & &  CXOGCJ174557.2-290301 &    266.48852 & -29.05042 & hard &      0.61 \\
SSTGC 0572559 & 266.48879 & -28.96801 &      9.52 &      9.09 &   \nodata &   \nodata & &  CXOGCJ174557.3-285804 &    266.48899 & -28.96787 & hard &      0.80 \\
SSTGC 0575612 & 266.49335 & -29.02236 &      8.91 &      8.90 &      8.62 &      8.92 & &  CXOGCJ174558.4-290120 &    266.49364 & -29.02232 & soft &      0.93 \\
SSTGC 0577500 & 266.49625 & -29.04868 &     10.02 &      9.83 &      9.44 &     10.04 & &  CXOGCJ174559.1-290255 &    266.49640 & -29.04875 & hard &      0.53 \\
SSTGC 0578722 & 266.49813 & -29.00052 &     10.09 &      9.93 &      9.74 &   \nodata & &  CXOGCJ174559.5-290002 &    266.49806 & -29.00067 & soft &      0.59 \\
SSTGC 0579221 & 266.49888 & -28.98583 &      9.61 &      9.60 &      9.26 &      9.22 & &  CXOGCJ174559.7-285908 &    266.49881 & -28.98571 & hard &      0.47 \\
SSTGC 0580810 & 266.50121 & -29.01808 &   \nodata &     11.52 &   \nodata &   \nodata & &  CXOGCJ174600.2-290105 &    266.50108 & -29.01809 & soft &      0.41 \\
SSTGC 0588034 & 266.51210 & -29.01352 &   \nodata &     10.80 &   \nodata &   \nodata & &  CXOGCJ174602.8-290049 &    266.51202 & -29.01366 & hard &      0.56 \\
\enddata
\tablecomments{Positional coincidence does not necessarily imply a physical association of IR and X--ray sources. In fact, $\sim70\%$ of the objects on this list are likely to be random positional coincidences of unrelated sources along the line of sight.}
\tablenotetext{a}{soft = detected in the 0.5--2.0 keV band; hard = not detected in the 0.5--2.0 keV band.}
\tablenotetext{b}{Separation between IR and X-ray source positions.}
\end{deluxetable}

\clearpage

\begin{figure}
\plotone{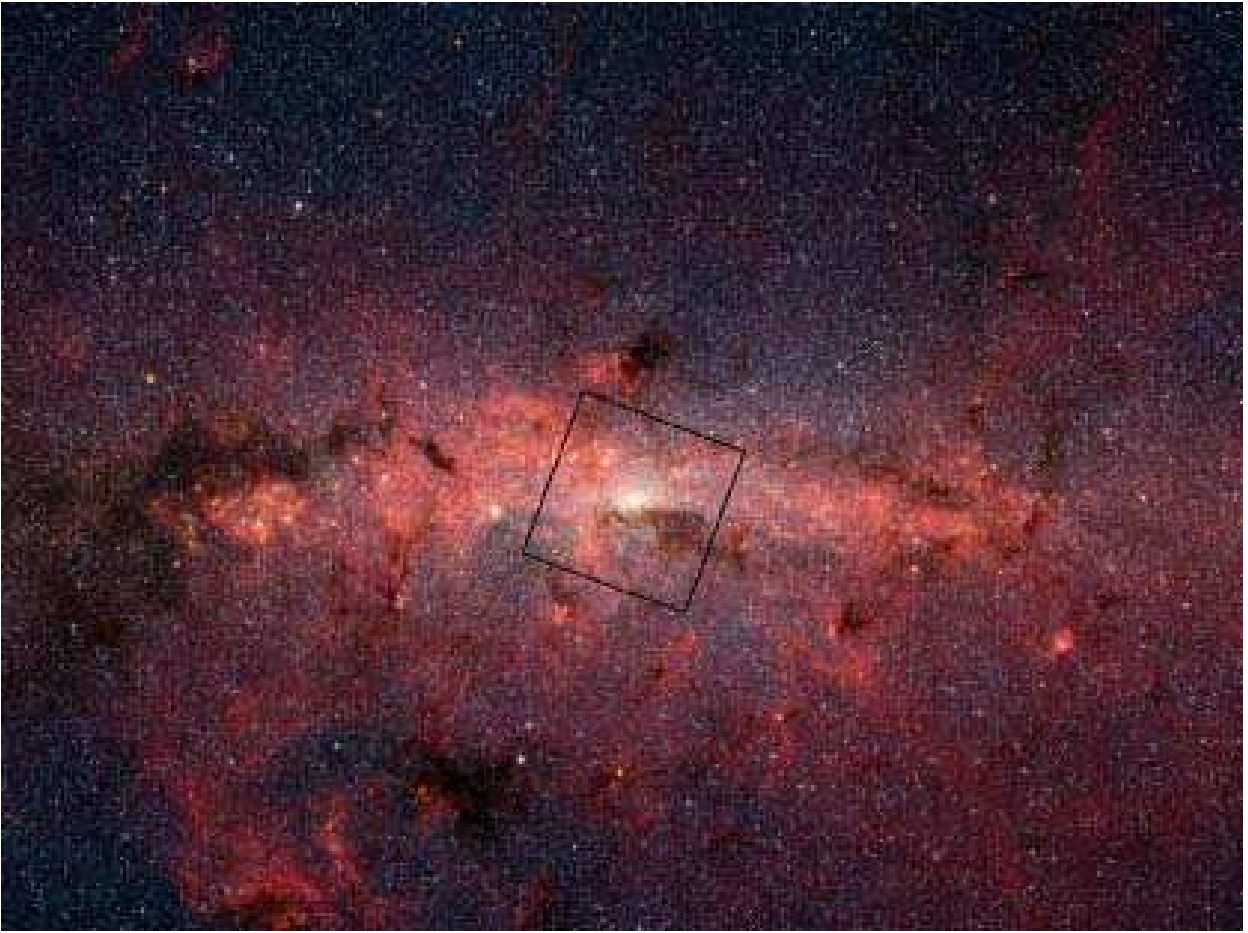}
\caption{Our {\it Spitzer}/IRAC Galactic Center Survey mosaic image (Stolovy et al.\ 2006), showing the $40 \times 40$ pc 
X--ray comparison field (square outline). This image is centered
at ($l$, $b$) = (0.0, 0.0) and is displayed in galactic
coordinates. \label{fig_1}}
\end{figure}

\begin{figure}
\plotone{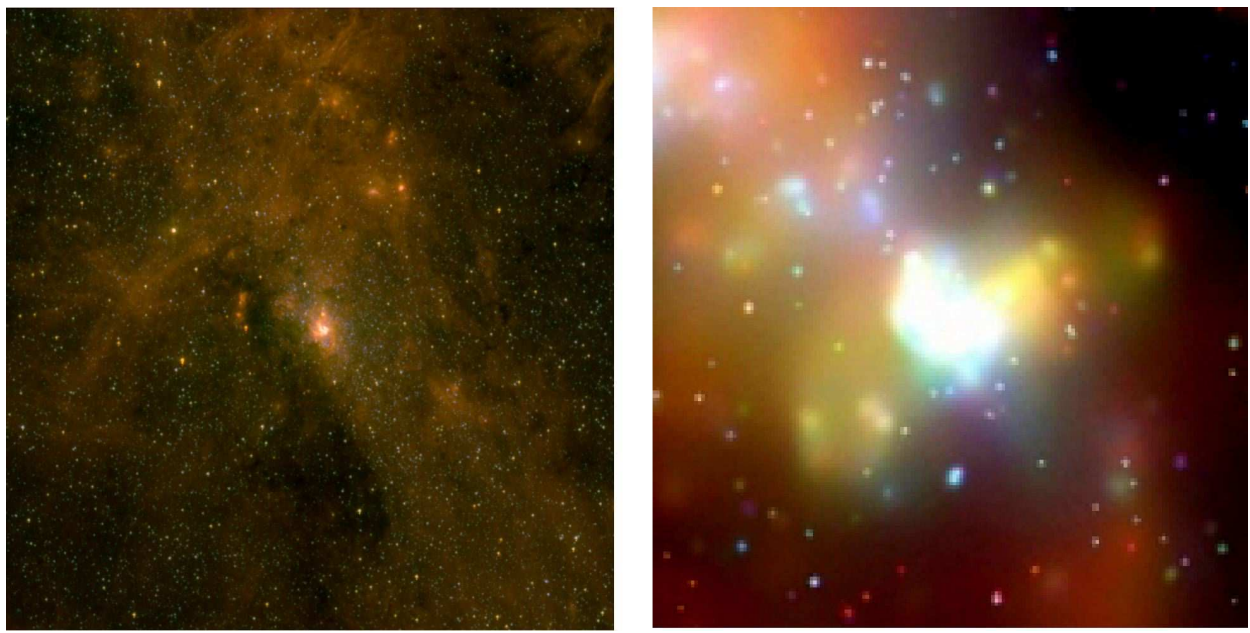}
\caption{The $40 \times 40$ parsec infrared and X--ray field that was 
used for the X--ray/infrared point source correlation study. {\it Left}: Composite 3.6 (blue), 5.8 (green) and 
8.0 $\micron$ (red) {\it Spitzer}/IRAC image of the $40 \times 40$ parsec study field. It yielded $\sim$20,000 
infrared point sources. {\it Right}: Composite Chandra image of the study field (Wang et al.\ 2002);  
red = 1 - 3 keV,  green = 3 - 5 keV, blue = 5 - 8 keV. Equatorial north is up, east is right. \label{fig_2}}
\end{figure}

\begin{figure} 
\plotone{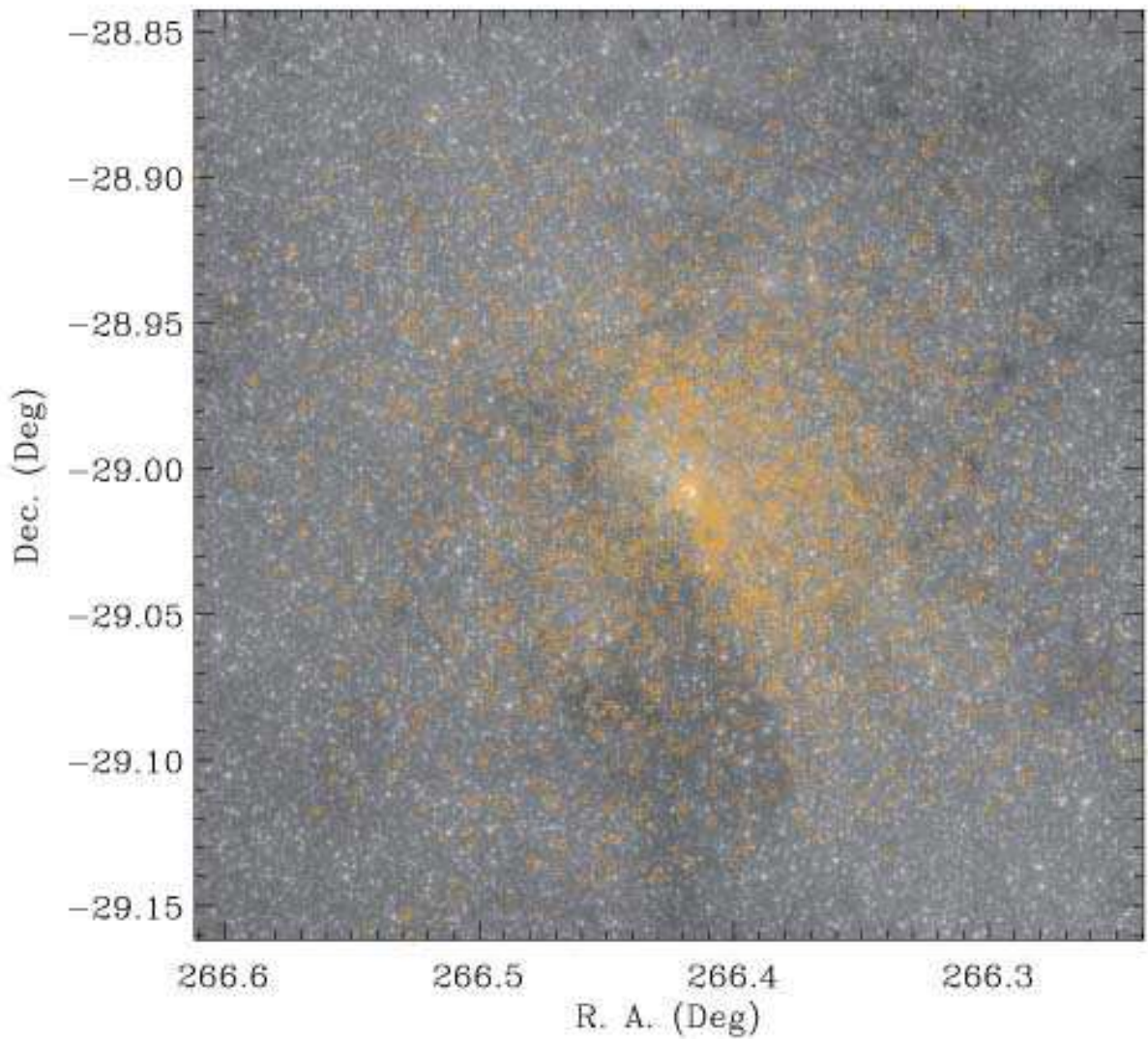} 
\caption{(a) Our {\it Spitzer}/IRAC 3.6 $\micron$ image, showing the $\sim$20,000 compact infrared sources brighter than $[3.6] = 12$ that we extracted in a $40 \times 40$ parsec subset of the survey region (Ram\'irez et al.\ 2008), overlaid with 2357 hard and soft X--ray sources (orange circles) identified by Muno et al.\ (2003). (b) The same 3.6 $\micron$ image is shown. Here, only the X--ray point sources that fell within $0\farcs5$ (blue), 1$''$ (green), and 2$''$ (red) of a 3.6 $\micron$ source are indicated. Hard X--ray sources are indicated by crosses; soft sources are indicated by circles. The six white squares indicate the locations of the massive young stars studied by Muno et al.\ (2006). All six are detected by IRAC, but only three are X--ray sources. \label{fig_ch1x}}
\end{figure} 
\clearpage 
{\plotone{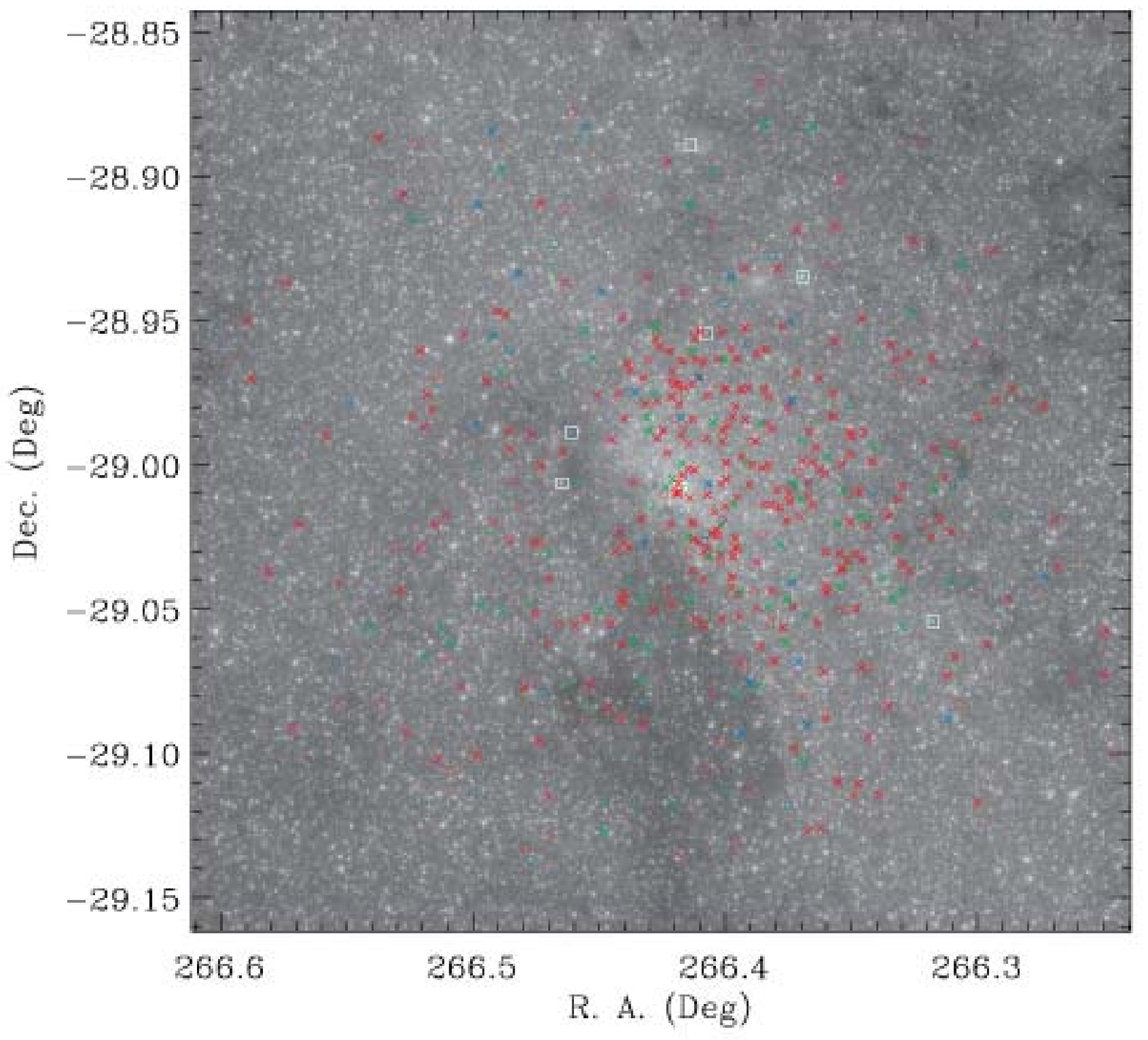}} 
\centerline{Fig. 3. --- Continued.} 
\clearpage 

\begin{figure} 
\plotone{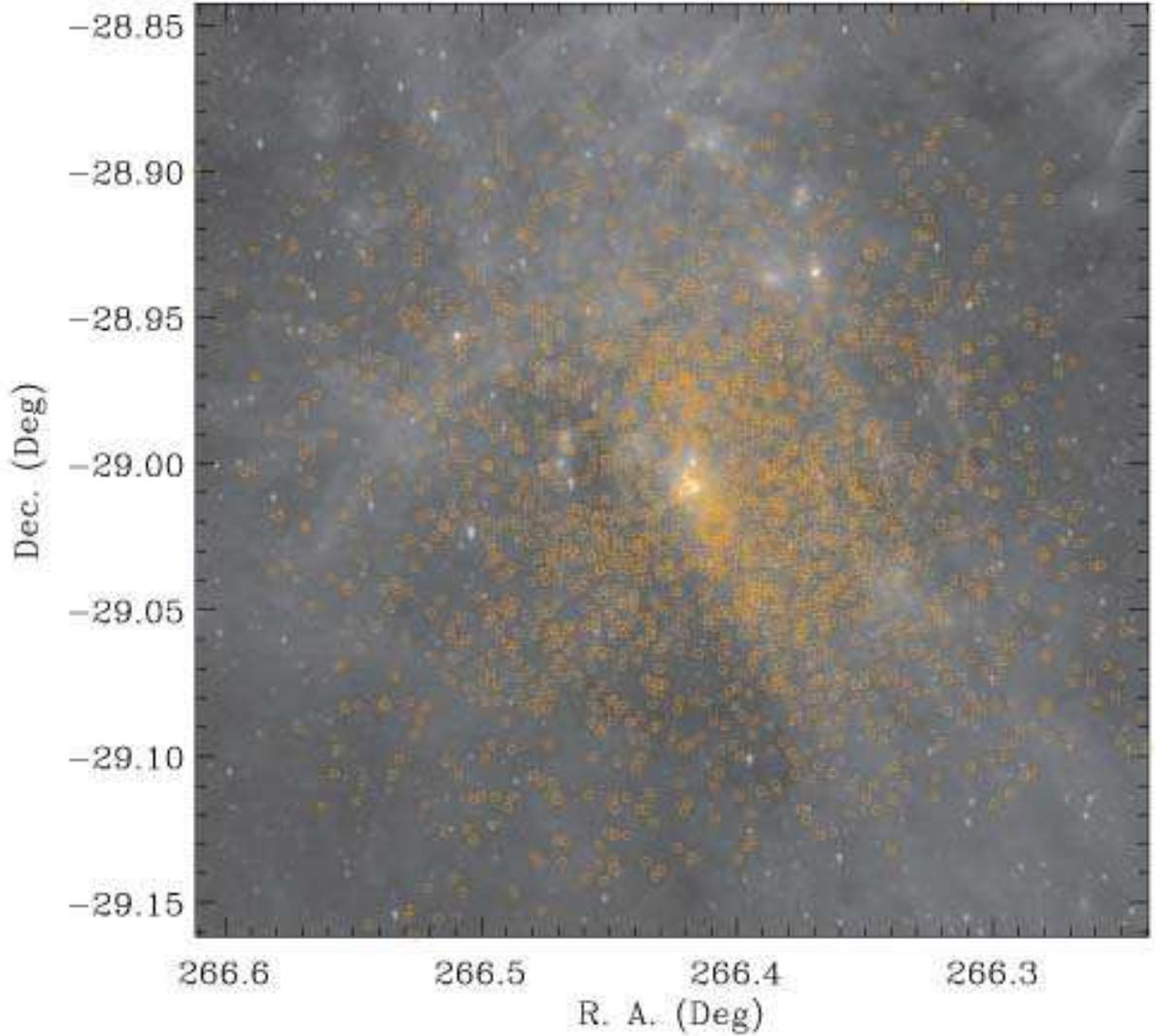} 
\caption{(a) Our {\it Spitzer}/IRAC 8.0 $\micron$ image, overlaid with the 2178 X--ray sources without IR counterparts (orange circles). (b) The same 8.0 $\micron$ image is shown. Here, only the X--ray point sources that fell within $0\farcs5$ (blue), 1$''$ (green), and 2$''$ (red) of an 8.0 $\micron$ source are indicated. Hard X--ray sources are indicated by crosses; soft sources are indicated by circles. The six white squares indicate the locations of the massive young stars studied by Muno et al.\ (2006). All six are detected by IRAC, but only three are X--ray sources. \label{fig_ch4x}}
\end{figure} 
\clearpage 
{\plotone{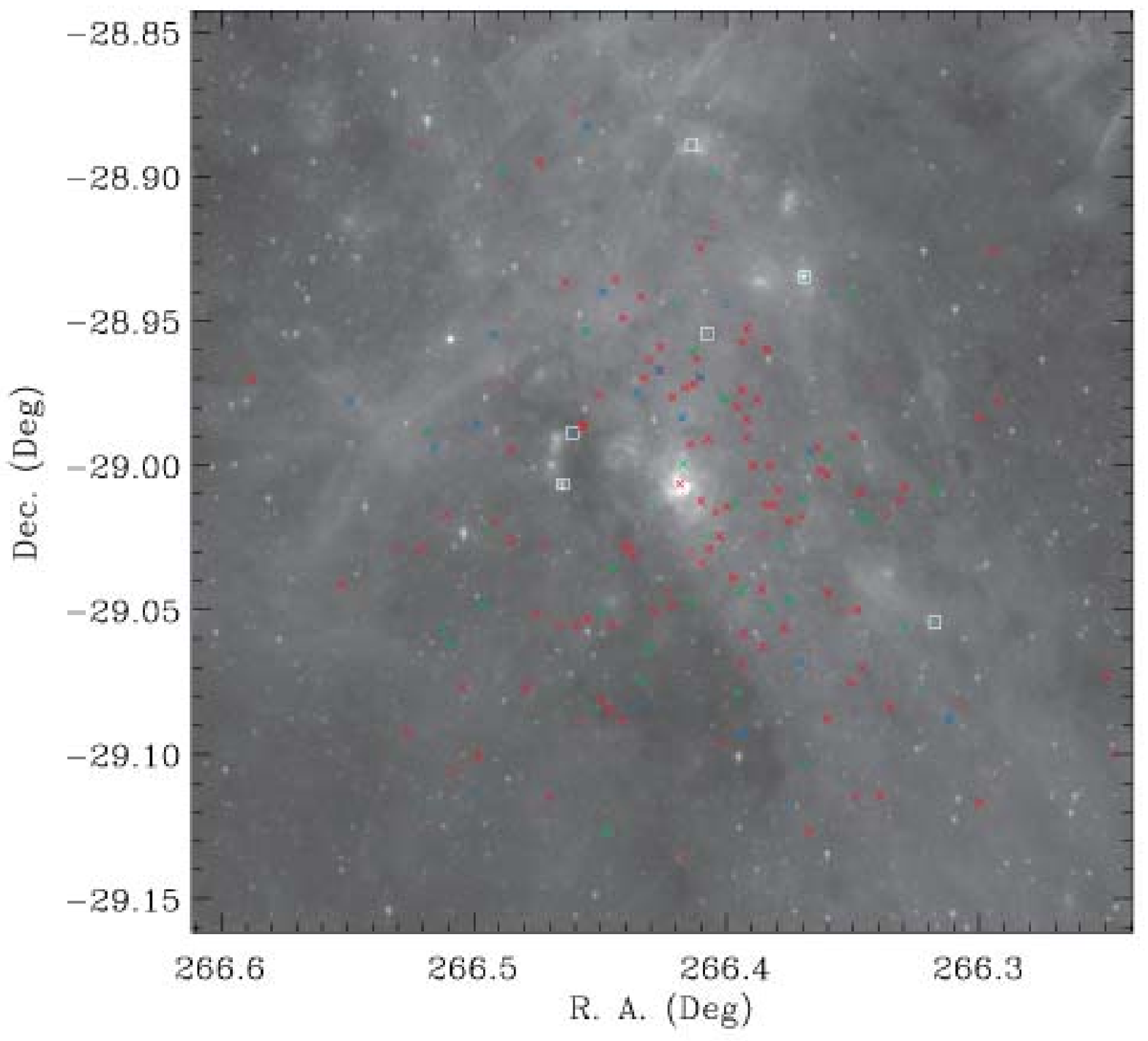}} 
\centerline{Fig. 4. --- Continued.} 
\clearpage

\begin{figure}
\plottwo{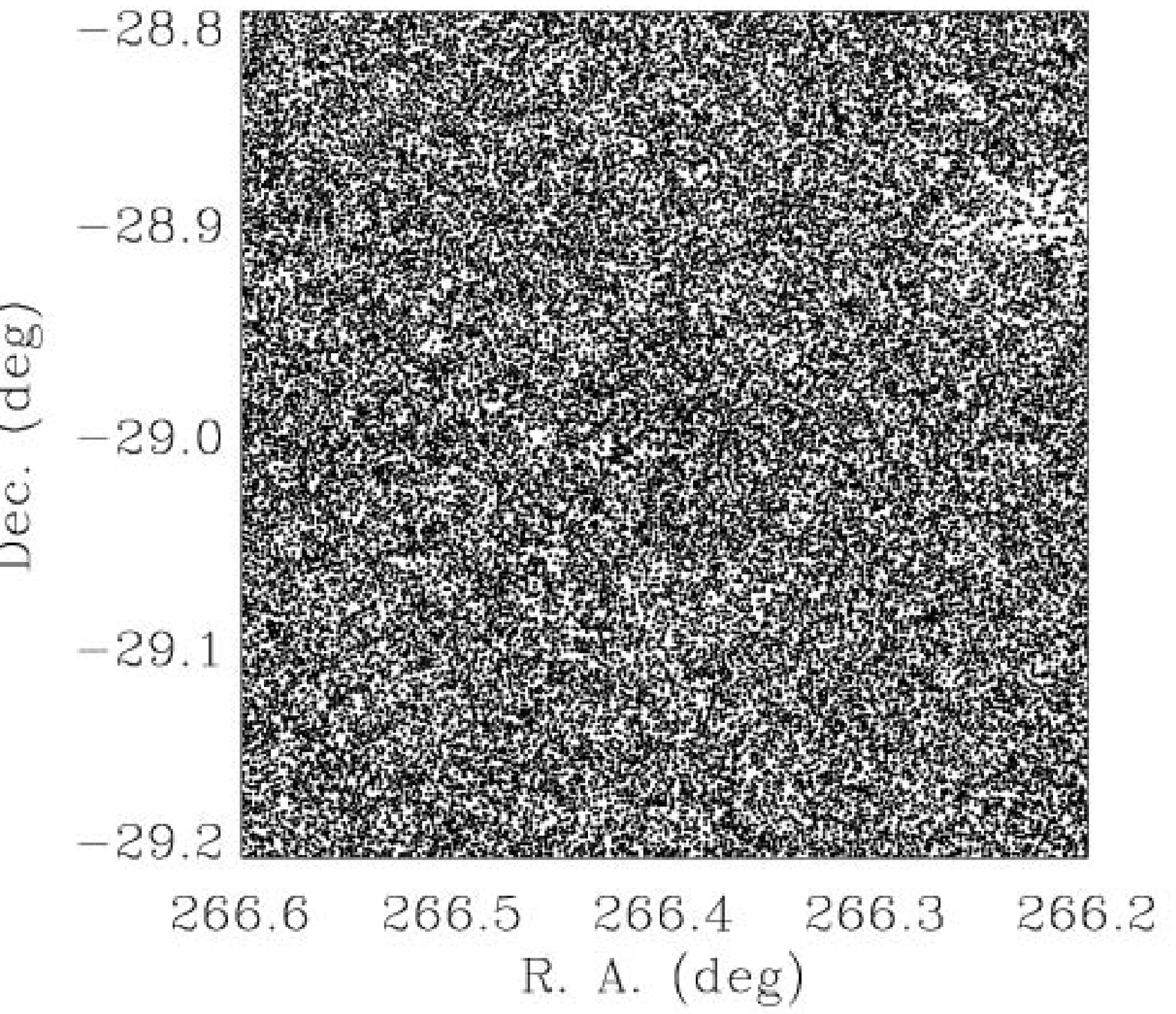}{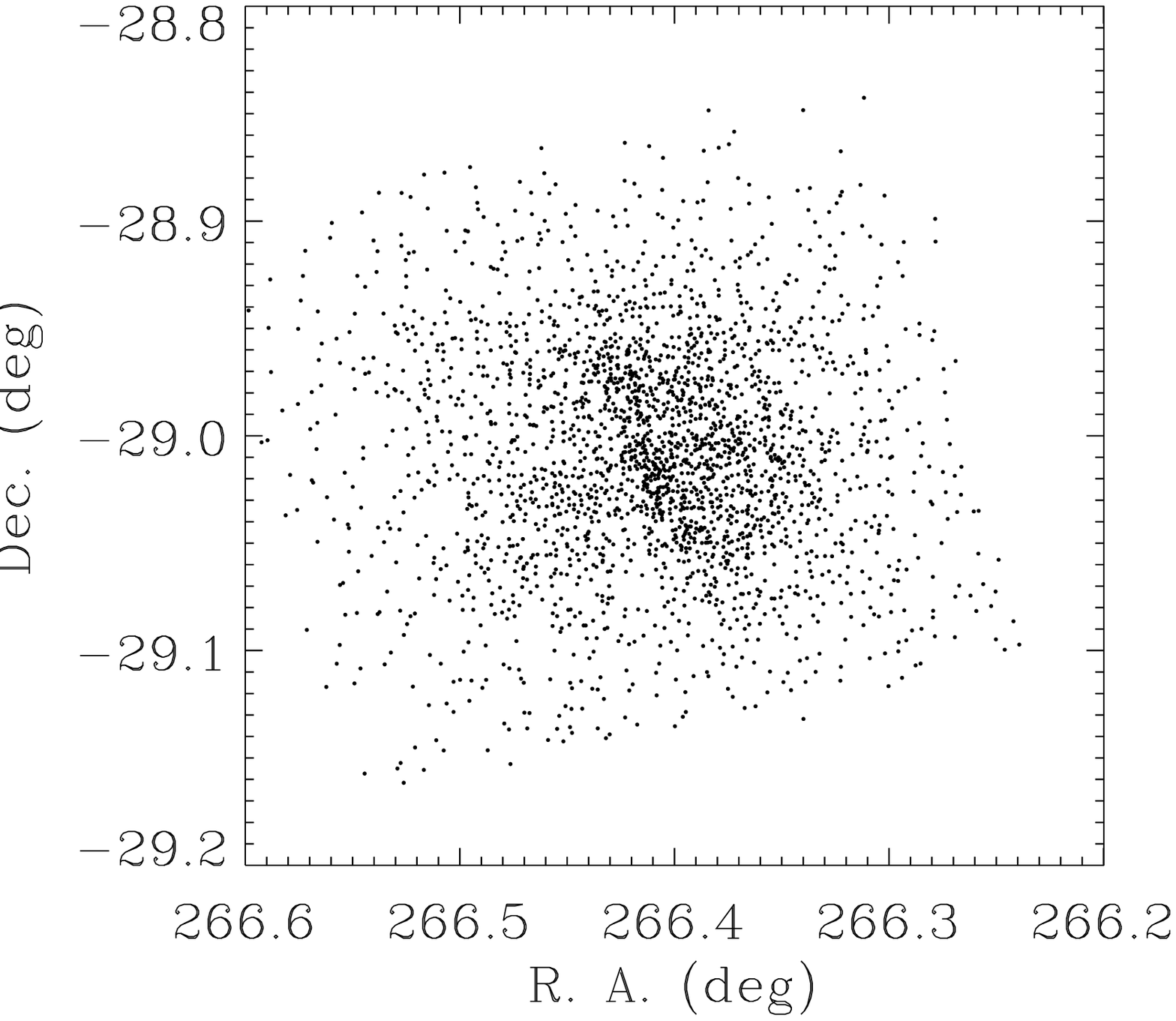}\\
\plottwo{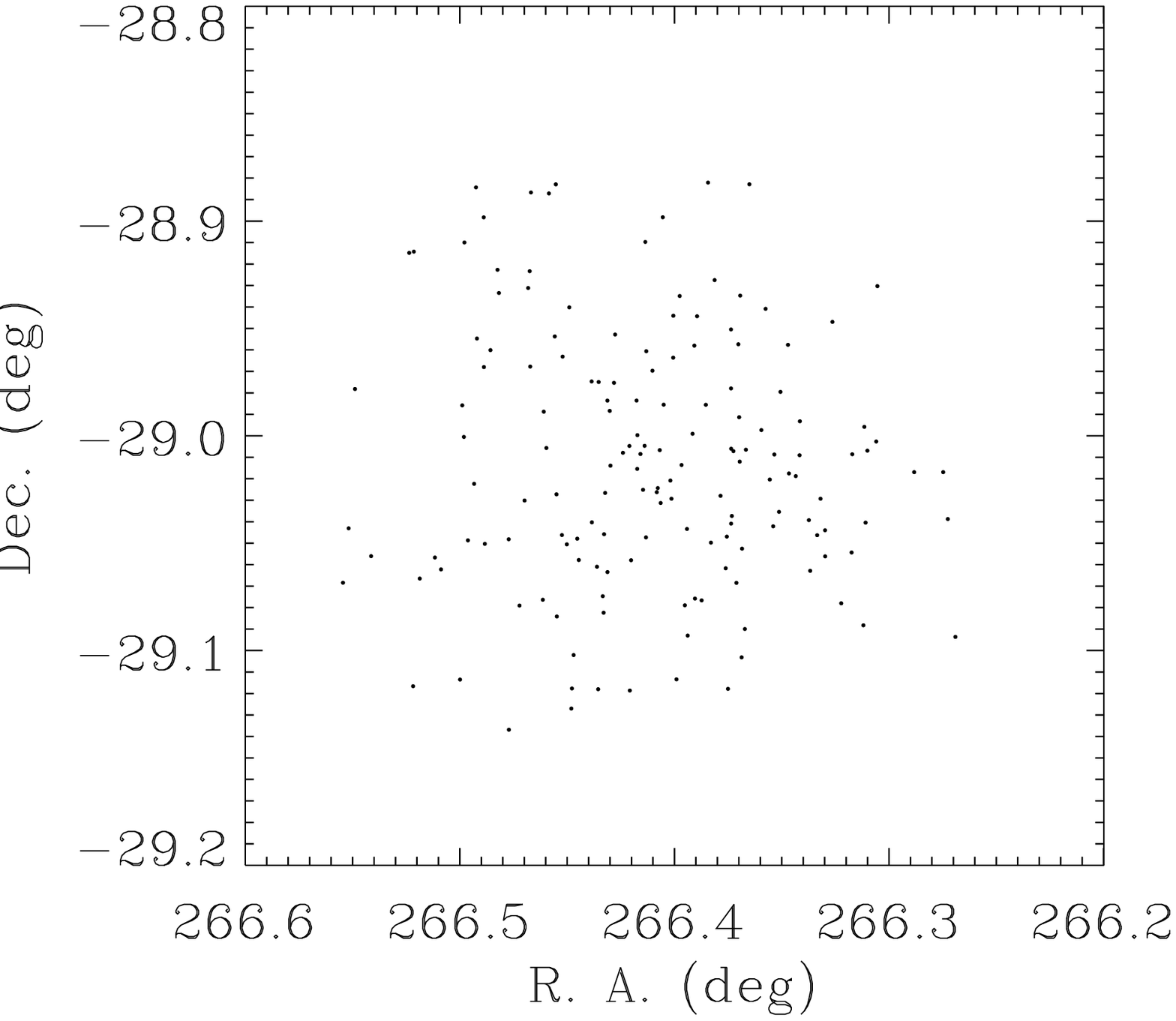}{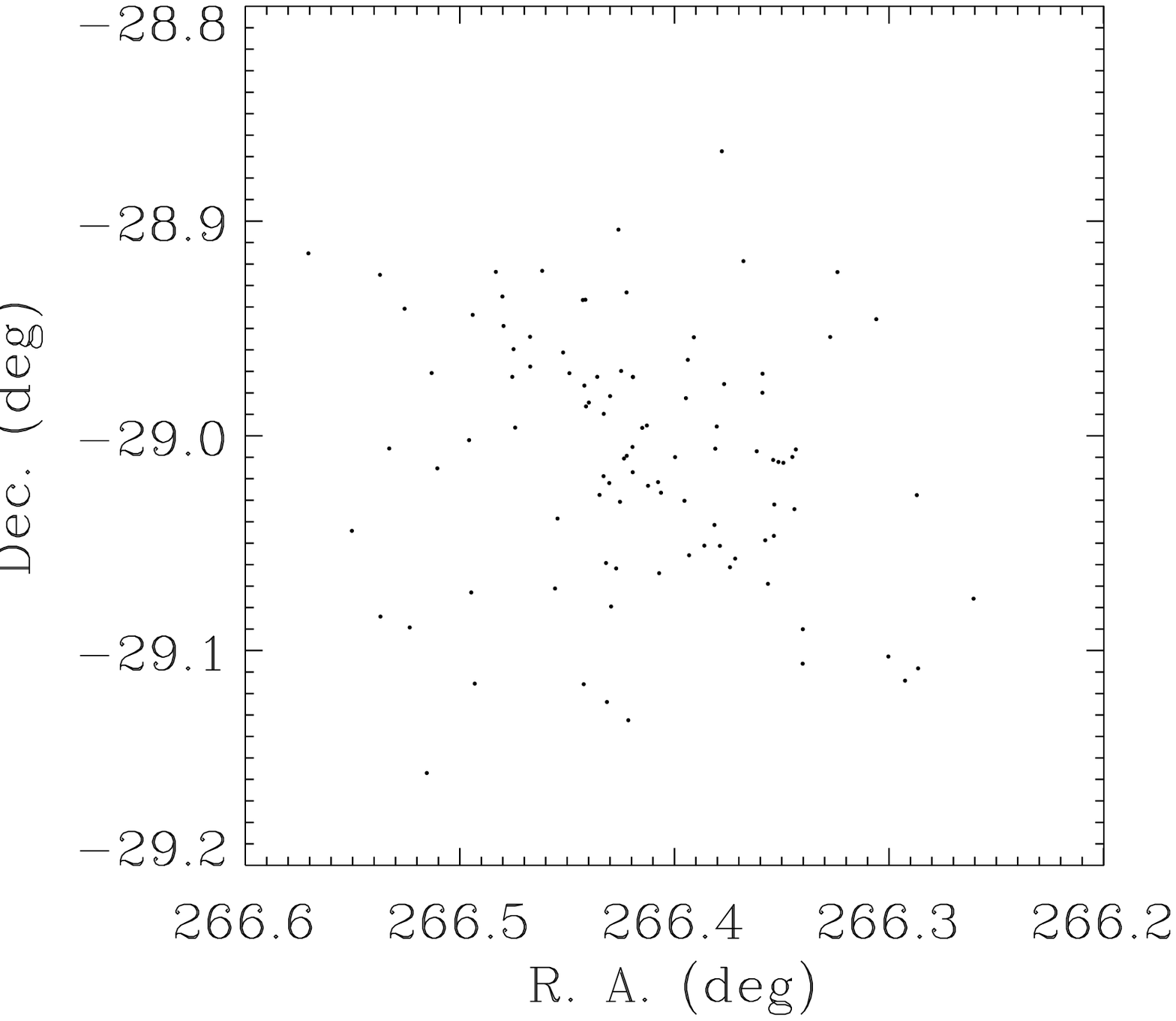}\\
\caption{(a) Positions of all $\sim$20,000 3.6 $\micron$ sources (black dots) in the $40 \times 40$ pc study field from our {\it Spitzer}/IRAC Galactic Center survey. 
(b) Positions of all 2357 X--ray sources in the Muno et al.\ (2003) survey.
(c) Positions of the 156 IRAC 3.6 $\micron$ sources that fall within 1$''$ of an X--ray source in the Muno et al.\ (2003) sample, a correlation of about 7\%.       
(d) Shifting the position template by 5$''$ from the source positions in (a) results in an average of 120 IRAC 3.6 $\micron$ sources within 1$''$ of  X--ray source, suggesting that these are chance associations. The actual correlation of 3.6 $\micron$ and X--ray point sources within 1$''$ is therefore probably not greater than 2\%.
\label{fig_dots}} 
\end{figure}

\begin{figure}
\plotone{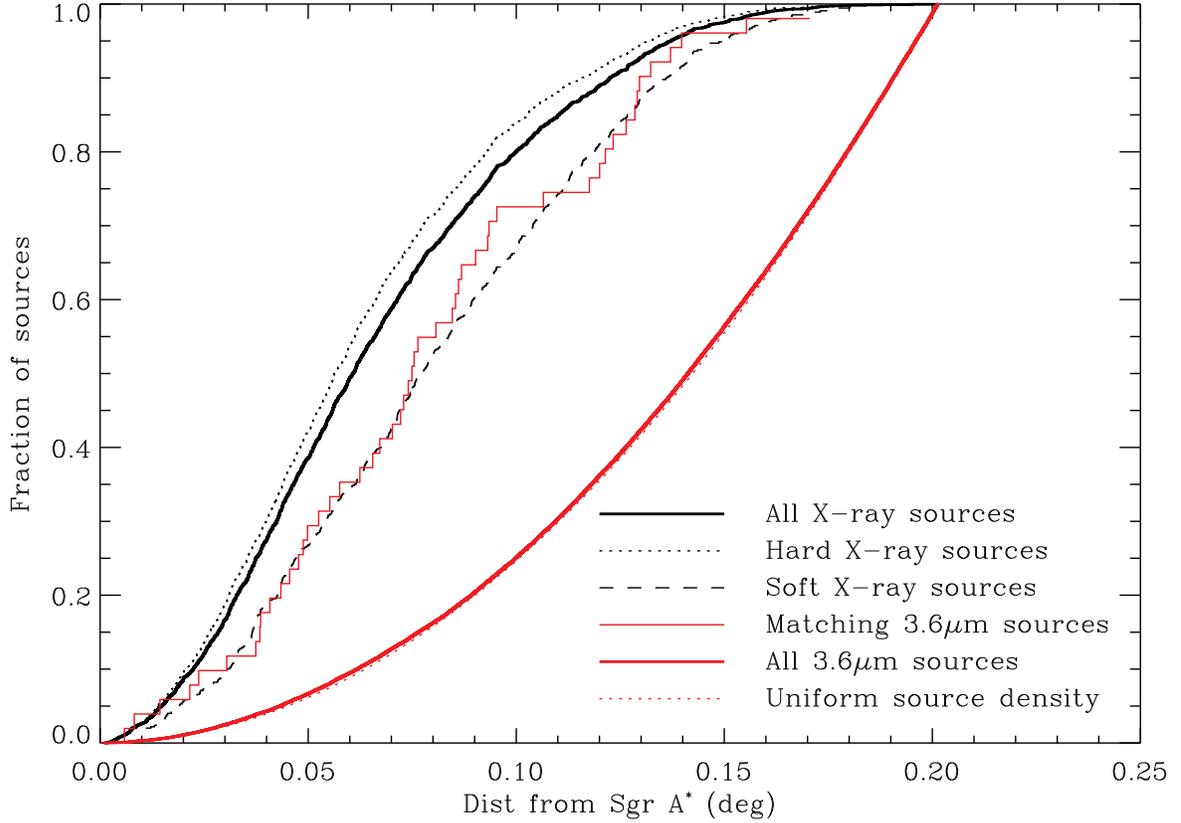}
\caption{The cumulative radial distributions with respect to Sgr A* are shown from 
all X--ray sources, only the hard sources, only the soft sources, and only 
the sources with matching 3.6 $\micron$ sources within $0\farcs5$.
The K--S test indicates that the hard sources have a statically different distribution
than the soft sources and the sources with IR counterparts. The distributions of 
soft sources and sources with 3.6 $\micron$ counterparts are {\it not} significantly different. The thick red line shows the corresponding distribution for all 3.6 $\micron$
sources within the same region as the X-ray sources. This distribution differs only slightly 
from that expected for a uniform source density across the region.
\label{fig_cumul}}
\end{figure}

\begin{figure}
\plotone{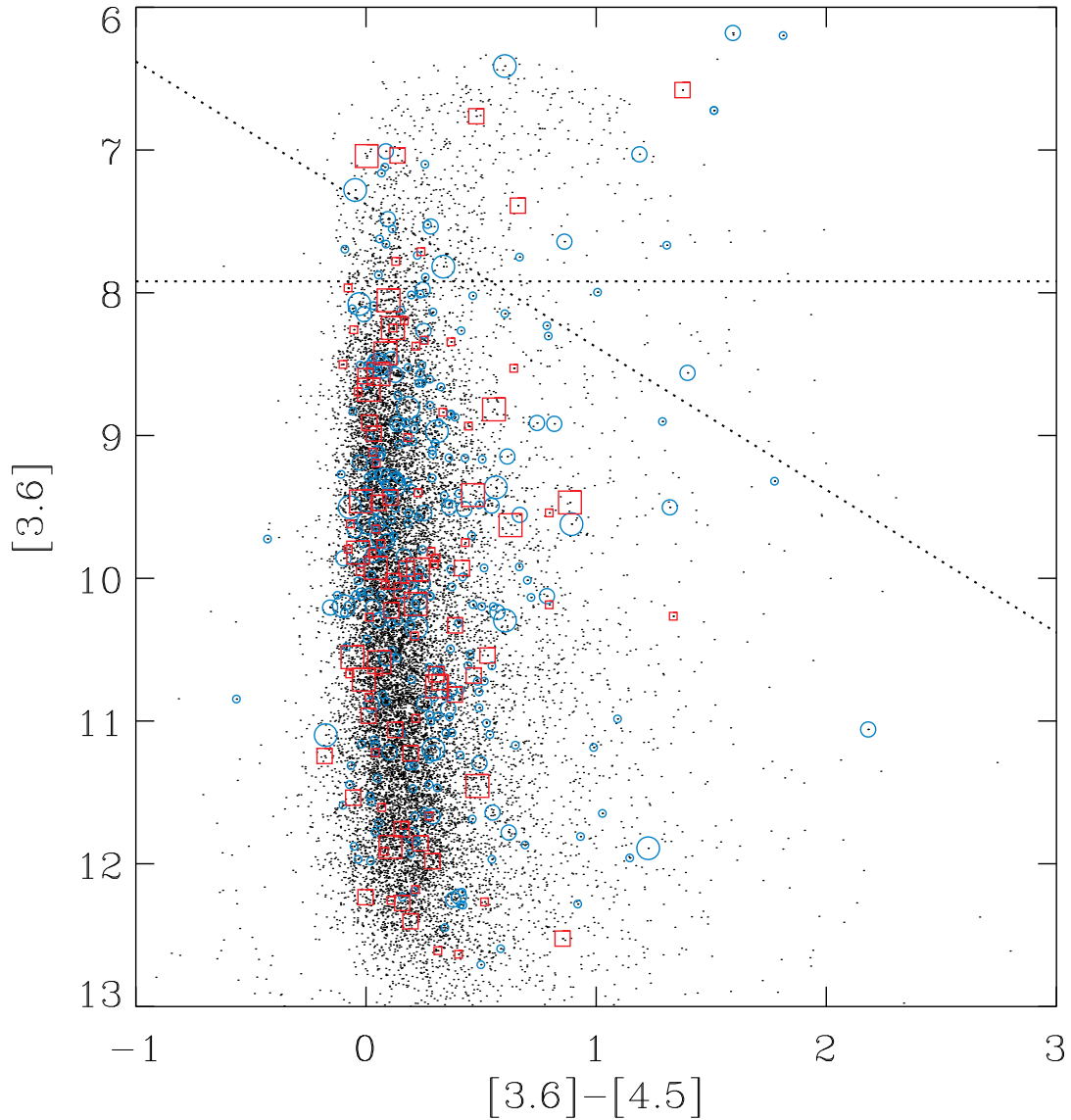}
\caption{IR stars with X--ray candidates have typical colors. Color-magnitude diagram for 17613 IRAC sources (small black dots) that were detected in both channels 1 and 2 (3.6 $\micron$ and 4.5 $\micron$) in the $40 \times 40$ parsec study region. 42 X--ray sources had IRAC sources lying within $0\farcs5$ of the nominal X--ray positions (large symbols), 130 had IRAC sources within 1$''$ (medium symbols), and 394 had IRAC sources within 2$''$ (small symbols). Square symbols indicate soft X--ray sources; circles indicate hard X--ray sources. The horizontal and diagonal dotted lines indicate the levels at which 
saturation may begin to affect the 3.6 and 4.5 $\micron$ photometry respectively (Ram\'irez et al.\ 2008). \label{fig_cmd}}
\end{figure}

\begin{figure}
\epsscale{0.6}
\plotone{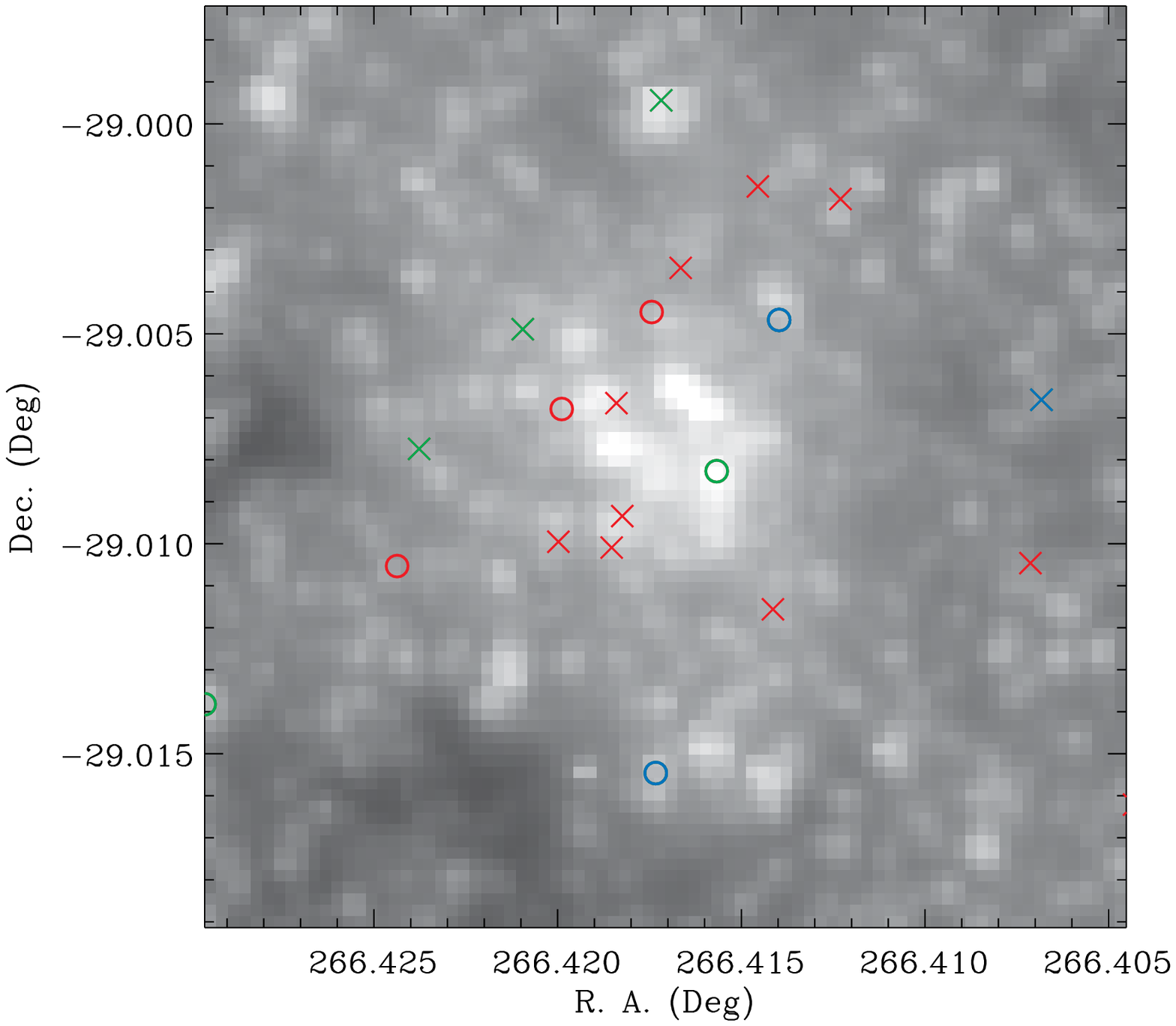}
\plotone{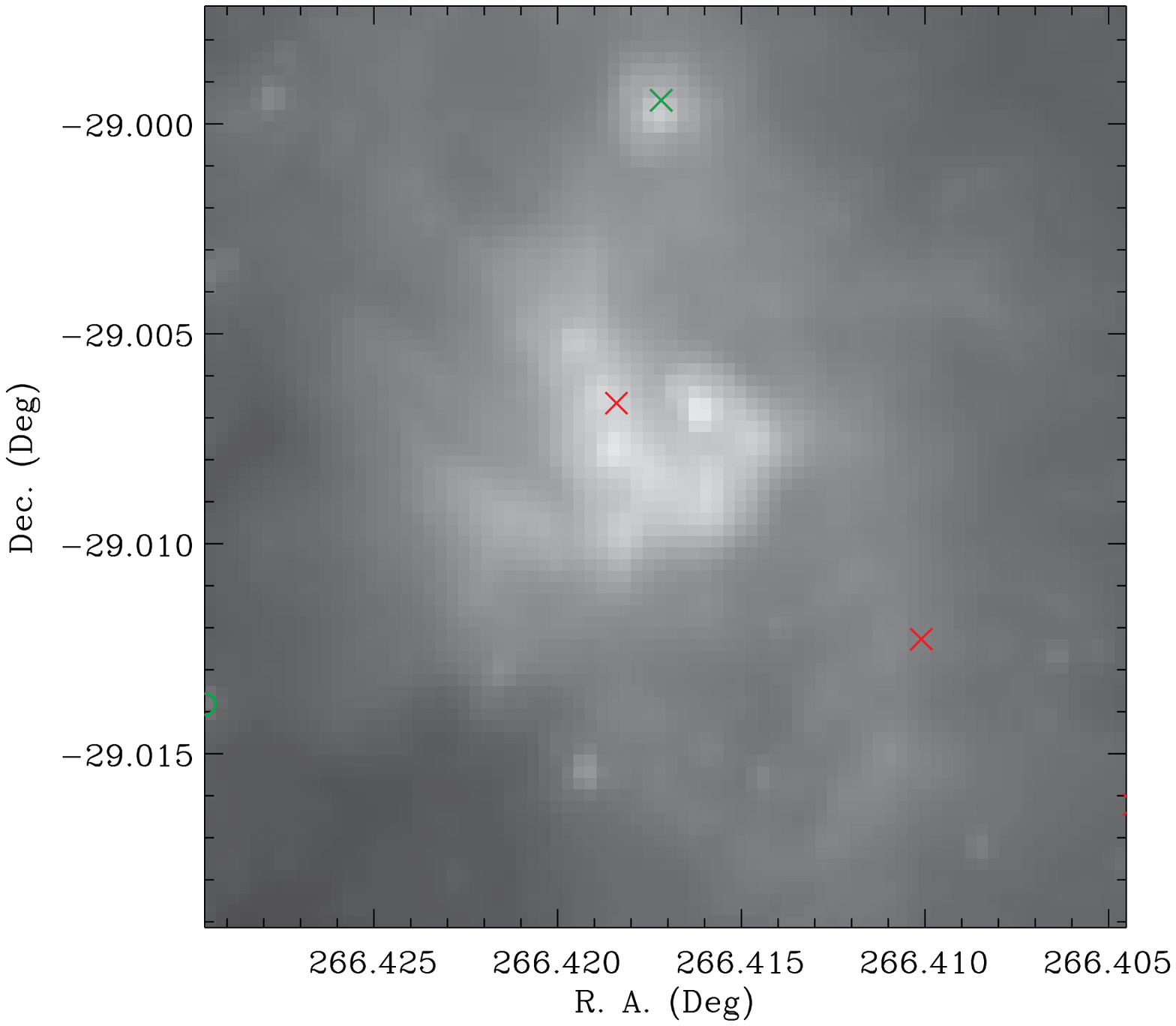}
\caption{(top) An enlarged view of matching X--ray and 3.6 $\micron$ point sources at 
the Galactic Center cluster. Matching sources within $0\farcs5$, 1$''$ and 2$''$ are 
indicated by blue, green and red symbols, respectively. Hard X--ray sources are indicated by crosses; soft sources are indicated by circles. IRS 13 is at the green circle 
nearest to the center of the image. (bottom) The same view at 
8 $\micron$ with the matching X--ray and 8 $\micron$ point sources superimposed.
\label{fig_zoom}}
\end{figure}


\begin{thebibliography}{}

\bibitem[Bandyopadhyay2005]{329} Bandyopadhyay, R. M., et al.\ 2005, \mnras, 364, 1195

\bibitem[Bandyopadhyay et al.(2006)]{2006smqw.confE..13B} Bandyopadhyay, 
R.~M., Gosling, A.~J., Blundell, K.~M., Podsiadlowski, P., Einkenberry, 
S.~A., Mikles, V.~J., Miller-Jones, J.~C.~A., \& Bauer, F.~E.\ 2006, VI 
Microquasar Workshop: Microquasars and Beyond 

\bibitem[Belczynski2004]{337} Belczynski, K., \& Taam, R. E. 2004, \apj, 616, 1159

\bibitem[Blum1996]{339} Blum, R. D., Sellgren, K., \& Depoy, D. L. 1996, \apj, 470, 864

\bibitem[Dean et al.(2005)]{2005A&A...443..485D} Dean, A.~J., et al.\ 2005, \aap, 443, 485 

\bibitem[Ebisawa et al.(2005)]{2005ApJ...635..214E} Ebisawa, K., et al.\ 
2005, \apj, 635, 214 

\bibitem[Flaherty et al.(2007)]{2007ApJ...663.1069F} Flaherty, K.~M., 
Pipher, J.~L., Megeath, S.~T., Winston, E.~M., Gutermuth, R.~A., Muzerolle, 
J., Allen, L.~E., \& Fazio, G.~G.\ 2007, \apj, 663, 1069 

\bibitem[Gezari1996]{349} Gezari, D., Dwek, E., \& Varosi, F., 1996, Proc. Symposium 169 of the IAU, 
	ed. L. Blitz \& P. Teuben, 231 

\bibitem[Gezari et al.(2006)]{2006JPhCS..54..171G} Gezari, D.~Y., et al.\ 
2006, Journal of Physics Conference Series, 54, 171 

\bibitem[Ghez2006]{356} Ghez, A., et al.\ 2005, \apj, 635, 1087

\bibitem[Indebetouw et al.(2005)]{2005ApJ...619..931I} Indebetouw, R., et 
al.\ 2005, \apj, 619, 931 

\bibitem[Kaplan et al.(2006)]{2006ApJ...649L.107K} Kaplan, D.~L., Moon, 
D.-S., \& Reach, W.~T.\ 2006, \apjl, 649, L107 

\bibitem[Laycock et al.(2005)]{2005ApJ...634L..53L} Laycock, S., Grindlay, 
J., van den Berg, M., Zhao, P., Hong, J., Koenig, X., Schlegel, E.~M., \& 
Persson, S.~E.\ 2005, \apjl, 634, L53 

\bibitem[Liu2006]{365} Liu, X.-W., \& Li, X.-D. 2006, \aap, 449, 135

\bibitem[Maillard2004]{367} Maillard, J. P., Paumard, T., Stolovy, S. R., Rigaut, F. 2004, \aap, 423, 155  

\bibitem[Mauerhan2007]{369} Mauerhan, J. C., Muno, M. P., \& Morris, M.  2007, \apj, 662, 574

\bibitem[Mikles2006]{371} Mikles, V. J., Eikenberry, S. S., Muno, M. P., Bandyopadhyay, R. M., \&
 Patel, S. 2006, \apj, 650, 203

\bibitem[Muno et al.(2003)]{2003ApJ...589..225M} Muno, M.~P., et al.\ 2003, 
\apj, 589, 225 

\bibitem[Muno et al.(2006)]{2006ApJ...638..183M} Muno, M.~P., Bower, G.~C., 
Burgasser, A.~J., Baganoff, F.~K., Morris, M.~R., \& Brandt, W.~N.\ 2006, 
\apj, 638, 183 

\bibitem[Peeples2007]{381} Peeples, M. S., Stanek, K. Z., \& Depoy, D. L. 2007, AcA, 57, 173

\bibitem[Pfahl2002]{383} Pfahl, E., Rappaport, S., \& Podsiadlowski, P.  2002, \apj, 571, L37

\bibitem[Press et al.(1986)]{1986nras.book.....P} Press, W.~H., Flannery, 
B.~P., Teukolsky, S.~A., \& Vetterling, W. T.\ 1986, Numerical Recipes: The Art of Scientific Computing, (Cambridge: Cambridge University Press)

\bibitem[Rahoui et al.(2008)]{2008arXiv:0802.2410} Rahoui, F., Chaty, S., 
Lagage, P.-O., \& Pantin, E.\ 2008, arXiv:0802.2410

\bibitem[Ramirez2008]{385} Ram\'irez, S. V., et al.\ 2008, \apjs, 175, 147

\bibitem[Ruiter2006]{391} Ruiter, A. J., Belczynski, K., \& Harrison, T. E.  2006, \apj, 640, L167 

\bibitem[Sidoli et al.(2001)]{2001A&A...368..835S} Sidoli, L., Belloni, T., 
\& Mereghetti, S.\ 2001, \aap, 368, 835 

\bibitem[Stolovy2007]{399} Stolovy S., et al.\ 2006, J. Phys. Conf. Series, 54, 176

\bibitem[Walter et al.(2006)]{2006A&A...453..133W} Walter, R., et al.\ 2006, \aap, 453, 133 

\bibitem[Wang et al.(2002)]{2002Natur.415..148W} Wang, Q.~D., Gotthelf, 
E.~V., \& Lang, C.~C.\ 2002, \nat, 415, 148 

\end{thebibliography}
\end{document}